# Satellite-Based Seasonal Fingerprinting of Methane Emissions from Canadian Dairy Farms Using Sentinel-5P

**Padmanabhan Jagannathan Prajesh[1,2], Kaliaperumal Ragunath[3], Miriam Gordon[1], and Suresh Neethirajan[1,4]***

[1] Faculty of Agriculture, Dalhousie University, Truro, Nova Scotia B2N 5E3, Canada
[2] Department of Remote Sensing and GIS, Tamil Nadu Agricultural University, Coimbatore 641003, India
[3] Center for Water and Geospatial Studies, Tamil Nadu Agricultural University, Coimbatore 641003, India
[4] Faculty of Computer Science, Dalhousie University, Halifax, Nova Scotia B2N 5E3, Canada
**\*** Correspondence: sneethir@gmail.com

**Abstract:** Methane ($CH_4$) emissions from dairy farming are a significant but under-quantified component of agricultural greenhouse gases. This study provides a satellite-based assessment of dairy-specific methane emissions across Canada using high-resolution Sentinel-5P TROPOMI data. By integrating spatial clustering of 1,701 dairy farms and processors, a quasi-experimental design with paired non-dairy reference regions, and seasonal pattern decomposition, we analyzed national and regional spatiotemporal emission trends. Results show persistently higher methane levels in dairy regions (mean difference: 16.99 ppb), with consistent fall-winter peaks. Notably, the dairy-specific methane anomaly, defined as the concentration difference between dairy and non-dairy regions declined by 62.25% from 2019 to 2024, with a sharp drop during 2022-2023 (-41.11%). Meanwhile, national methane levels rose by 3.83%, with increasing spatial heterogeneity across provinces. An inverse relationship between baseline methane levels and growth rates suggests a convergence effect. Seasonal analysis revealed universal spring minima and fall-winter maxima, offering distinct temporal signatures for source attribution. This study demonstrates the value of satellite-based monitoring for policy-relevant methane assessments and introduces a scalable framework applicable to other regions. The observed narrowing of dairy methane anomaly indicates evolving emission dynamics, potentially reflecting rising baseline methane rather than a definitive reduction in dairy source emissions. This highlights the need for integrated satellite and ground-based approaches to enhance understanding and guide mitigation efforts.

## 1. Introduction

Atmospheric methane ($CH_4$) has emerged as a critical focus in climate change mitigation due to its potent warming potential of 28 - 34 times that of carbon dioxide over a 100-year timeframe [1]. Global atmospheric methane concentrations have risen from approximately 722 parts per billion (ppb) in pre-industrial times to over 1900 ppb recently, with accelerated growth since 2007 [2]. This trajectory significantly impacts Paris Agreement climate targets, as methane contributes approximately 0.5°C to observed global warming [3]. Agricultural activities represent a substantial and potentially manageable emission source among various contributors [4]. The dairy sector contributes significantly to agricultural methane emissions through enteric fermentation and manure management [5].

In Canada, dairy production accounts for approximately 8% of agricultural greenhouse gas emissions, primarily methane [6]. With approximately 977,000 dairy cows across 10,095 farms, the Canadian dairy industry presents both challenges and opportunities for targeted emission reduction [7]. However, effective mitigation requires robust spatiotemporal characterization of methane emissions at regional scales data historically limited by ground-based measurement constraints [8]. Traditional quantification approaches have relied on bottom-up inventory methods, applying emission factors to activity data such as livestock populations and management practices [9]. While providing valuable baseline estimates, these methods often fail to capture complex spatial and temporal dynamics influenced by regional climate variations, management practices, and ecosystem interactions [10]. Significant discrepancies between inventory estimates and atmospheric observations exist, as documented by [11], who found Western Canadian energy operation methane emissions nearly twice those reported in official inventories. Similar discrepancies may exist in agricultural emission estimates, highlighting the need for independent verification methods.

The Sentinel-5 Precursor (Sentinel-5P) satellite with its TROPOspheric Monitoring Instrument (TROPOMI) has revolutionized atmospheric methane monitoring with unprecedented spatial and temporal resolution [12]. With $5.5 \times 7$ km² spatial resolution and daily global coverage, TROPOMI measurements enable detection of regional methane enhancements and their temporal evolution, creating opportunities for top-down verification of emission estimates and identification of mitigation targets [12]. These observations provide column-averaged dry-air mole fractions of methane ($XCH_4$) sensitive to surface-level emissions, facilitating attribution to specific sources or regions [13]. While satellite-based monitoring offers significant advantages, attributing observed concentration patterns to specific sources remains challenging due to atmospheric transport, varying background concentrations, and spatial overlap of multiple emission sources [14]. These challenges are pronounced in agricultural regions, where emissions may coincide with wetlands, fossil fuel infrastructure, or other sources [15]. Addressing these attribution challenges requires robust analytical frameworks to isolate agricultural signals from background variations and competing sources.

Previous studies have attempted to quantify agricultural methane emissions using satellite observations through inverse modeling [16], enhancement detection over known source regions [12], and correlation with activity data [9]. However, these approaches have largely focused on continental or global scales, with limited application to regional agricultural systems like Canada's dairy sector. Few studies have characterized the seasonal dynamics of agricultural methane emissions using satellite observations with a critical gap given emission processes' sensitivity to seasonal environmental factors and management practices [17]. The Canadian dairy sector presents an interesting case study for satellite-based methane monitoring due to several factors: (1) geographical distribution spanning diverse climatic zones, creating natural experimental conditions for examining environmental modulation of emission patterns; (2) relatively standardized management practices governed by supply management systems, potentially reducing confounding influences observed in other countries [18]; and (3) northern latitude providing unique conditions for examining seasonal emission patterns due to pronounced temperature variations, frozen soil periods, and seasonal changes in cattle housing and feeding regimes.

We have developed a multi-method approach to characterize the spatiotemporal distribution of methane concentrations in Canadian dairy regions and identify seasonal emission patterns. Our approach combines satellite remote sensing data with spatial clustering techniques, quasi-experimental design elements, and advanced time series analysis to address attribution challenges. By comparing dairy-intensive regions with carefully selected non-dairy reference areas, we aim to isolate the dairy-specific methane signal while controlling for regional and latitudinal background variations. This study addresses four primary research questions focused on: (1) spatiotemporal dynamics of atmospheric methane concentrations across Canadian dairy regions (2019-2024); (2) comparison of methane concentrations between dairy-intensive and non-dairy reference areas; (3) seasonal patterns characterizing methane emissions from Canadian dairy regions; and (4) identification of consistent seasonal patterns with subtle regional variations that could support source attribution efforts when combined with other methane source types, and inform targeted mitigation strategies.. These insights will contribute to research utilizing advanced remote sensing technologies to improve understanding of agricultural greenhouse gas emissions and support evidence-based climate policy development.

## 2. Materials and Methods

The study uses a multi-method approach to assess methane ($CH_4$) emissions from dairy farming in Canada, leveraging satellite remote sensing data. The main objectives are to characterize the spatiotemporal distribution of methane concentrations and identify weekly composite based seasonal emission patterns. A quasi-experimental design (Figure 1) was applied, with dairy regions as treatment areas and non-dairy regions as controls, facilitating a comparative analysis while controlling for regional and latitudinal differences. This approach allows for a detailed understanding of methane emissions from dairy operations under varying environmental conditions.

*2.1. Data Acquisition and Processing*
2.1.1. Satellite Platform and Senor Specification

Atmospheric methane observations were derived from the Sentinel-5 Precursor (Sentinel-5P) satellite, which carries the TROPOspheric Monitoring Instrument (TROPOMI). Launched in October 2017, Sentinel-5P operates in a near-polar, sun-synchronous orbit with a 13:30 local solar time equatorial crossing. TROPOMI, a nadir viewing spectrometer with a 2600 km swath width, provides daily global coverage by measuring reflected solar radiation across ultraviolet (267–332 nm), ultraviolet-visible (305–499 nm), near-infrared (661–786 nm), and shortwave infrared (SWIR; 2300–2389 nm) spectral bands [19]. The SWIR measurements are critical for retrieving column-averaged dry-air mole fractions of methane ($XCH_4$) at a spatial resolution of $5.5 \times 7$ km². Given its sensitivity to surface-level methane concentrations, $XCH_4$ data are well-suited for detecting and characterizing localized emission sources.

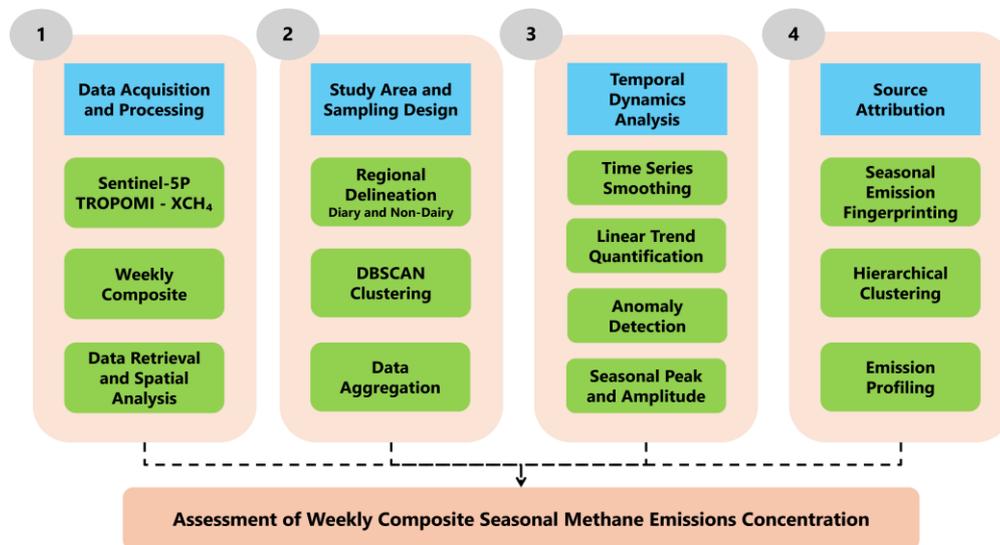

**Figure 1.** Systematic Representation of Characterizing Seasonal Variability of Methane Emissions from Dairy-Intensive Landscapes Using Weekly Data Composit

2.1.2. Data Reterival and Temporal Agrregation

Sentinel-5P Level 3 product were accessed via Google Earth Engine (GEE) API. The dataset comprising the XCH$_4$ column averaged dry air mixing ratio expressed in parts per billion (ppb) was retrieved for the period from 2019 - 2024, to investigate seasonal variabilities. The Sentinel-5P XCH$_4$ Level 3 products are pre-processed datasets that incorporate geolocation corrections, radiometric calibration, cloud masking, and bias corrections addressing albedo effects, aerosol interference, and systematic instrumental biases to ensure high data fidelity [19]. To balance temporal resolution and data quality, data were processed into weekly composites, minimizing gaps from cloud interference and instrumental errors. Spatial filtering isolated methane concentrations in dairy-intensive agricultural zones for comparative analysis with non-dairy regions. Quality control excluded data points with cloud fractions more than 20% and instrumental anomalies through quality assurance flags. This workflow ensures that we process methane data at both high temporal and spatial resolution, enabling comprehensive analyses of seasonal and regional methane emissions.

*2.2. Study Area Definfiatin and Sampling Stratergy*

2.2.1. Dairy Region Delineation and Reference Region Selection

To identify concentrated dairy farming areas, we applied the DBSCAN (Density-Based Spatial Clustering of Applications with Noise) spatial clustering method [20,21] on a geospatial database comprising 1,701 geolocated dairy farms and associated processing facilities across ten Canadian provinces (Figure 1). These locations were retrieved through a systematic Google Maps-based geocoding process, enabling spatially explicit analysis at a national scale.

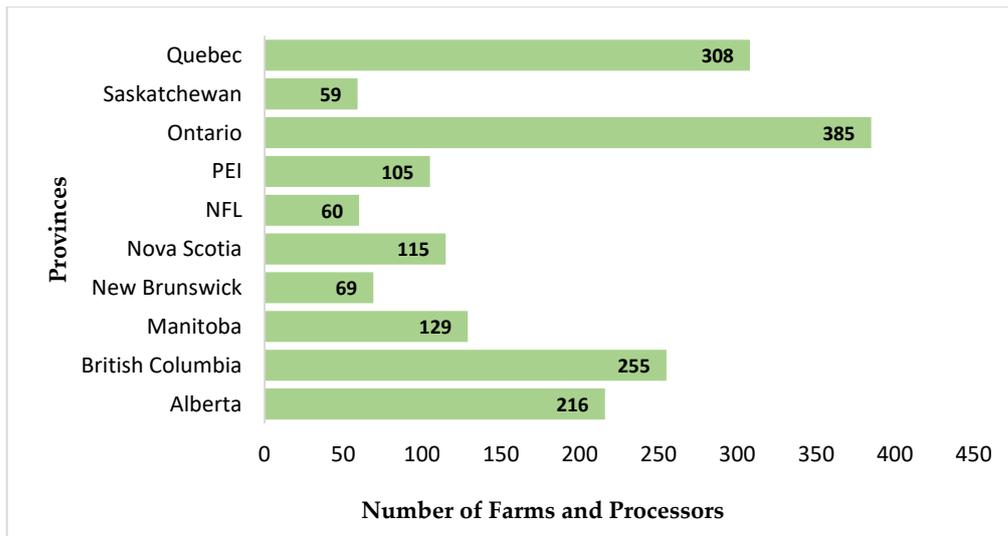

**Figure 2.** Distribution of Dairy Farms and Processors across the provinces in Canada. PEI - Prince Edward Island, NFL - Newfoundland and Labrador

To delineate regions of concentrated dairy activity, we employed bounding box geometries at approximately 2° × 2° in size, following a spatial clustering to maintain methodological consistency. Fourteen rectangular bounding boxes were defined to encompass high-density dairy farming zones (Table 1). These regions recognized by Dairy Farmers of Canada (2023), represent around 82% of Canada's dairy output [22], making them ideal for analyzing dairy-related methane fluxes. To support comparative analysis, we identified eight reference regions situated at similar longitudes but at higher latitudes relative to the dairy zones with minimal livestock activity (Table 1).

To minimize potential confounding factors, reference regions were selected to avoid major anthropogenic and natural methane sources, including urban areas, wetlands, and industrial zones. Selection was informed by cross-referencing the ESRI Sentinel-2 Land Use/Land Cover (S2 LULC) dataset availbale at https://livingatlas.arcgis.com/landcover/, enabling identification of areas predominantly characterized by natural vegetation or barren land. This ensured a clean atmospheric background for robust attribution of observed methane enhancements to dairy-related activities. This spatial pairing strategy adheres to Intergovernmental Panel on Climate Change [23] guidelines for atmospheric methane monitoring, ensuring the maintenance of comparable climatic regimes while minimizing the influence of emissions.

**Table 1.** Canadian dairy and reference regions based on bounding box coordinates

| Province | Region | West | South | East | North |
|---|---|---|---|---|---|
| **Dairy Regions** | | | | | |
| British Columbia | Fraser Valley | 123.0°W | 49.0°N | 121.5°W | 49.5°N |
| British Columbia | Vancouver Island | 125.3°W | 48.3°N | 123.4°W | 50.0°N |
| Alberta | Central | 114.5°W | 51.0°N | 112.5°W | 53.0°N |
| Alberta | Southern | 114.0°W | 49.5°N | 112.0°W | 51.0°N |
| Saskatchewan | Central | 107.0°W | 50.5°N | 104.0°W | 52.5°N |
| Manitoba | Interlake | 98.0°W | 50.0°N | 96.0°W | 52.0°N |
| Manitoba | Eastern | 97.5°W | 49.0°N | 95.5°W | 50.5°N |
| Ontario | Southern | 83.0°W | 42.0°N | 79.0°W | 44.5°N |
| Ontario | Eastern | 78.0°W | 43.5°N | 74.5°W | 45.5°N |
| Quebec | St Lawrence Valley | 73.0°W | 45.0°N | 70.0°W | 47.5°N |
| Quebec | Eastern Township | 72.5°W | 45.0°N | 70.5°W | 46.0°N |
| New Brunswick | Central | 67.5°W | 45.5°N | 65.5°W | 47.0°N |
| Nova Scotia | Annapolis Valley | 65.5°W | 44.5°N | 64.0°W | 45.5°N |
| Prince Edward Island | Central | 63.5°W | 46.0°N | 62.5°W | 46.5°N |
| **Reference (Non-Dairy) Regions** | | | | | |
| British Columbia | Northern | 118.0°W | 56.0°N | 114.0°W | 58.0°N |
| Alberta | Northern | 118.0°W | 56.0°N | 114.0°W | 58.0°N |
| Saskatchewan | Northern | 108.0°W | 55.0°N | 104.0°W | 57.0°N |
| Manitoba | Northern | 99.0°W | 55.0°N | 95.0°W | 57.0°N |
| Ontario | Northern | 90.0°W | 50.0°N | 86.0°W | 53.0°N |
| Quebec | Northern | 75.0°W | 50.0°N | 70.0°W | 53.0°N |
| New Brunswick | Northern | 68.0°W | 47.0°N | 66.0°W | 48.0°N |
| Nova Scotia | Cape Breton | 61.5°W | 45.5°N | 60.0°W | 47.0°N |

The overall methodological framework draws on established approaches in satellite-based greenhouse gas monitoring *[24]* and addresses spatial heterogeneity issues *[25]*. The implementation of this regionalization strategy improves the signal-to-noise ratio in detecting methane emissions linked to dairy production across varied Canadian ecological and climatic zones. By controlling regional variability and background methane concentrations, this approach provides a robust foundation for attributing observed methane anomalies to dairy-related sources with improved spatial and temporal fidelity.

2.2.2. Spatial Clustering Using DBSCAN Algorithm

To refine the initially defined bounding boxes based on actual farm distributions, we applied the Density-Based Spatial Clustering of Applications with Noise (DBSCAN) algorithm, a widely used unsupervised machine learning method for spatial pattern recognition [26]. This technique enabled the identification of natural clusters of dairy farms based on their geographic proximity, thereby improving the representativeness of the defined dairy regions.

The epsilon (eps) value was set to 0.1, which corresponds approximately to a 10 km radius at mid-latitudes in Canada. This value defines the maximum distance between two farms for them to be considered part of the same cluster. The minimum sample parameter was set to 2, indicating the minimum number of dairy farms required to form a dense cluster. This clustering approach was applied to the DBSCAN algorithm, which has been successfully used for identifying spatial patterns in geocoded datasets [27–29]. The DBSCAN algorithm defines the cluster as follows (Eq. 1):

$$C = \{ x \in X \mid \exists p \in X: x \in N\varepsilon(p) \land | N\varepsilon(p) | \geq Min_{Pts} \} \quad \ldots (1)$$

where:
- $C$ is the set of all points belonging to a cluster,
- $X$ is the complete dataset of farm locations,

- $N\varepsilon(p)$ denotes the ε neighborhood of a point p, and
- $Min_{Pts}$ is the minimum number of neighbors within ε required to form a cluster.

Dairy region bounding boxes were defined a priori based on expert knowledge of Canada's dairy production geography. Spatial clustering algorhtim: DBSCAN was then applied within these regions to identify dense farm clusters, allowing for the exclusion of outliers and improving the spatial precision of methane emission analyses. This clustering process allowed us to delineate core dairy zones more accurately and to exclude isolated farms or noise points from regional statistical analysis. The refined spatial delineation improves both the spatial precision and interpretability of our subsequent methane emission analyses.

2.2.3. Regional Boundary Refinement Using Adaptive Bounding Box

Following the identification of dairy farm clusters, we implemented an adaptive bounding box refinement technique to define spatial extents more precisely around each cluster. This method calculated the geographic boundaries by identifying the minimum and maximum latitudinal and longitudinal coordinates for each cluster, then extending these boundaries with a fixed spatial buffer. The initial bounding region $R$ is expressed as (Eq. 2):

$$R = [L_{min} - \beta, L_{max} + \beta] \times [\lambda_{min} - \beta, \lambda_{max} + \beta] \quad \dots (2)$$

where:
- $L_{min}$ and $L_{max}$ represents the minimum and maximum latitudes,
- $\lambda_{min}$ and $\lambda_{max}$ represents the minimum and maximum longitudes of the spatial points within a cluster, and
- $\beta$ is used as a user defined buffer parameter.

This approach provided sufficient spatial granularity while preserving region-level comparability. This approach was informed by [24], who developed dynamic targeting techniques for satellite observations, addressing the limitations of fixed boundaries in emission source detection. Our buffer implementation followed [30], who emphasized the use of adaptive spatial margins to reduce retrieval uncertainties and prevent underestimation of emissions at boundary pixels.

*2.3. Spatial Analytical Implementation*

2.3.1. Zonal Statistics Computation

We computed zonal statistics for each defined region by aggregating pixel-level methane concentrations using Google Earth Engine's reducer functions. The mean and standard deviation of methane values within each region $R$ were calculated (Eq. 3) and (Eq. 4).

$$\mu_R = \frac{1}{|R|} \sum_{p \epsilon R} v(p) \quad \dots (3)$$

$$\sigma_R = \sqrt{\frac{1}{|R|} \sum_{p \epsilon R} (v(p) - \mu_R)^2} \quad \dots (4)$$

where:
- $v(p)$ represents the methane concentration at pixel p, and
- $|R|$ denotes the number of valid pixels within the region $R$.

To balance computational efficiency and spatial resolution, we selected a scale parameter of 1000 meters, which aligns with best practices in large-scale satellite-based atmospheric trace gas retrieval [31]. The zonal summary statistics were extracted using Earth Engine's reduceRegion() functions applied to TROPOMI derived methane imagery. This approach allows for region-specific aggregation of satellite observations while maintaining the spatial fidelity needed for subnational emissions assessments. We implemented the computation using the following Python function which evaluates spatially averaged concentrations for each region geometry, enabling downstream statistical comparisons between dairy and non-dairy zones.

2.3.2. Spectral Reflectance Modelling and Correction of Atmospheric Methane

We generated continuous spectral surfaces of atmospheric methane concentrations by compositing satellite observations over time using the arithmetic means of individual image acquisitions. To ensure consistency across spatial analyses, we exported the resulting mean image with a standardized spatial resolution of 5 km and a coordinate reference system (CRS) of EPSG:4326. The export was performed using the Earth Engine export function. We also applied a two-dimensional Gaussian filter to the composite surfaces to suppress high-frequency noise while preserving key spatial gradients in methane concentrations. The Gaussian smoothing kernel is defined as (Eq. 5):

$$G(x, y) = \frac{1}{2\pi\sigma^2} e^{-\frac{x^2+y^2}{2\sigma^2}} \quad \dots (5)$$

where, $\sigma = 1.0$ was selected based on empirical assessments of spatial smoothness and edge preservation. This denoising technique is widely used in atmospheric remote sensing to enhance the interpretability of trace gas concentration maps [32].

2.3.3. National and Provincial Scale Data Aggregation of Methane Metrics

To assess methane dynamics at broader administrative levels, regional estimates were aggregated to provincial and national scales using area-weighted averaging. This approach accounts for spatial heterogeneity in region sizes, thereby ensuring accurate representation of methane concentrations across varying administrative extents [12]. At the provincial level, the average methane concentration $\mu_P$ for each province $p$ was computed as (Eq. 6):

$$\mu_p = \frac{\sum_{r \epsilon p} A_r \times \mu_r}{\sum_{r \epsilon p} A_r} \quad \dots (6)$$

where, $\mu_r$ is the mean methane concentration for region $r$ within province $p$, and $A_r$ is the area of region $r$. Subsequently, the national average methane concentration $\mu_n$ was calculated by aggregating the provincial estimates using a similar area-weighted approach (Eq. 7):

$$\mu_n = \frac{\sum_{p=1}^{P} A_p \times \mu_p}{\sum_{p=1}^{P} A_p} \dots (7)$$

where, $A_P$ represents the area of province $p$, and $P$ is the total number of provinces. This hierarchical aggregation framework facilitates scalable methane monitoring while preserving spatial fidelity at multiple administrative levels.

*2.4. Temporal Dynamics and Time Series Analysis*

2.4.1. Kalman Filtering for Time Series Smoothing

To reduce noise while preserving meaningful trends in the time series data, a Kalman filter was implemented. The Kalman filter is a recursive algorithm used to estimate the state of a system by combining observations, models, and their associated uncertainties. The filter aims to minimize the mean squared error between predicted and actual states [33]. It operates in two main stages: the "prediction" step, where it generates estimates of the current state and their uncertainties, and the "update" step, where these predictions are refined by incorporating new observed measurements, which may contain errors and noise. During the update step, a weighted average is applied, giving more importance to estimates with higher confidence. The prediction (Eq. 8-9) and update (Eq. 10-12) steps of the filter are expressed through the following equations:

Prediction Step:

$$\hat{x}_{t|t-1} = \hat{x}_{t-1|t-1} \quad \ldots (8)$$
$$P_{t|t-1} = P_{t-1|t-1} + Q \quad \ldots (9)$$

Update Step:

$$K_t = \frac{P_{t|t-1}}{P_{t|t-1} + R} \quad \ldots (10)$$

$$\hat{x}_{t|t} = \hat{x}_{t|t-1} + K_t \left(z_t - \hat{x}_{t|t-1}\right) \quad \ldots (11)$$

$$P_{t|t} = (1 - K_t) P_{t|t-1} \quad \ldots (12)$$

where:
- $\hat{x}_{t|t-1}$ represents a priori state estimate,
- $P_{t|t-1}$ is a priori estimate covariance,
- $K_t$ is the Kalman gain,
- $z_t$ is the measurement at time t,
- $Q$ is the process noise covariance (set to 0.01), and
- $R$ is the measurement noise covariance (set to 0.1).

For this analysis, the noise covariance process was set to $Q = 0.01$ and the measurement noise covariance to $R = 0.1$ was based on optimization through empirical performance. To fine tune the filter parameters, we employed a grid search optimization technique, minimizing the root mean square error (RMSE) between raw and filtered values. This approach ensures the optimal selection of parameters for filtering atmospheric methane data with minimal error [34]. The Kalman filter is particularly effective for linear systems with Gaussian noise, as it reduces high-frequency fluctuations while preserving the underlying trends and seasonal patterns, which are essential for accurate time series environmental monitoring [35].

2.4.2. Non-Parametric and Parametric Trend Estimation

We estimated linear trends in atmospheric methane ($CH_4$) concentrations to quantify the annual rate of change across the study regions. The trend was computed using the difference between final and initial mean methane values over time, normalized to provide an annualized rate [16,36] based on (Eq. 13):

$$Trend = \left(\frac{CH_{4(final)} - CH_{4(initial)}}{\Delta t}\right) \times 365 \quad \ldots (13)$$

where:

- $CH_{4(final)}$ and $CH_{4(initial)}$ represent the final and initial mean methane concentrations respectively, and
- $\Delta t$ is the duration of the observation period in days.

To assess statistical significance, we applied the non-parametric Mann-Kendall trend test, which is robust against missing data and non-normal distributions typical in satellite-derived time series [37]. This approach aligns with best practices in methane inversion studies and environmental monitoring

2.4.3. Statistical Analysis of Methane Anomaly Time Series

To quantitatively assess the convergence pattern between dairy and non-dairy regions, we conducted time series analysis on the weekly methane concentration differential, denoted as Δt, defined as (Eq. 14):

$$\Delta t = CH_4(dairy, t) - CH_4(reference, t) \quad ...(14)$$

where, $CH_4(dairy, t)$ and $CH_4(reference, t)$ represents the mean methane concentration in the dairy and non dairy regions respectively at week $t$. We then evaluated the temporal trend in $\Delta t$ using linear regression (Eq. 15):

$$\Delta t = \beta_0 + \beta_1 t + \epsilon_t \quad ...(15)$$

where, $\beta_0$ is the intercept, $\beta_1 t$ is the slope representing the rate of change in methane differential (ppb/week), $t$ is the time (in weeks) and $\epsilon_t$ is the error term. Additionally, we employed the non-parametric Mann-Kendall test to assess trend significance without assuming normality, suitable for environmental time series with potential outliers. Confidence intervals (95%) for the slope were computed, and the total change over the study period was calculated in both absolute (ppb) and percentage terms, providing robust statistical evidence for the observed convergence between dairy and non-dairy methane concentrations.

2.4.3. Inter-Regional Convergence Analysis

To quantify spatial convergence in methane concentrations over time, a modified sigma-convergence framework was employed, following the approach of [38]. Specifically, the coefficient of variation (CV) was used as a normalized measure of dispersion across regions at each time point $t$ (Eq. 16):

$$CV_t = \frac{\sigma_t}{\mu_t} \quad ...(16)$$

where, $\sigma_t$ represents the standard deviation of regional methane concentrations, and $\mu_t$ is the corresponding mean across all regions at time $t$. The temporal trends in regional convergence were evaluated by computing the negative time derivative of the CV, defined as (Eq. 17):

$$Convergence\ Rate = -\frac{d(CV_t)}{dt} \quad ...(17)$$

A positive convergence rate indicates decreasing spatial disparities in methane concentrations, signifying a trend toward inter-regional homogenization. This dynamic metric enables systematic assessment of spatial equity in emission patterns over time.

2.4.4. Methane Anomaly Detection and Quantification

We quantified dairy-specific methane ($CH_4$) anomalies by calculating the difference between average concentrations in dairy regions and their corresponding reference (non-dairy) regions at each weekely time step. In this study the dairy specific methane anomaly ($\Delta CH_4$) is defined as the excess column averaged methane concentration in dairy region relative to reference non-dairy region. The anomaly was computer at time $t$ (Eq. 18):

$$\Delta CH_4(t) = \frac{1}{n_d}\sum_{i=1}^{n_d} CH_{4_{d,i,t}} - \frac{1}{n_r}\sum_{j=1}^{n_r} CH_{4_{r,j,t}} \quad \dots (18)$$

Where,

- $CH_{4_{d,i,t}}$ represents the methane concentration in dairy region $i$ at time $t$,

- $CH_{4_{r,j,t}}$ denotes the methane concentration in reference region $j$ at the same time, and

- $n_d$ and $n_r$ are the total number of dairy and reference regions, respectively.

To identify periods of anomalously elevated emissions, we defined a significant methane anomaly as any weekly value of $\Delta CH_4$ exceeding two standard deviations above the mean of the full anomaly time series. This threshold was determined using a statistically grounded approach based on the distribution of weekly anomaly values over the entire study period (2019–2024), comprising 314 weekly observations.

Based ont this threhold, the significanct anomaly frequency (SAF) was computed annually thus representing the percentage of weeks with statistically signficant anomalies (Eq. 19):

$$SAF\ (\%) = \left(\frac{N_{\Delta CH_4} > Mean\ Anomaly\ Threshold}{N_{total}}\right) \times 100 \quad \dots (19)$$

Where,

- $N_{\Delta CH_4} > Mean\ Anomaly\ Threshold$ is the number of weeks which the methane anomaly exceeds the threshold, and
- $N_{total}$ is the total number of weeks in the year.

This approach controls for shared background variability by referencing non-dairy regions with similar environmental conditions, thus isolating potential methane enhancements linked to dairy production. By leveraging spatiotemporal differentials, this method builds upon established practices in methane source attribution [16,24] , offering a robust framework for detecting localized agricultural emissions within broader atmospheric signals.

2.5. Seasonal Dynamics and Pattern Recognition

2.5.1. Seasonal Decomposition from Methane Time Series

We segmented the annual cycle into four meteorological seasons [39] consistent with Northern Hemisphere conventions: winter (December - February), spring (March - May), summer (June - August), and fall (September - November). For each region and season, we computed the following summary statistics for atmospheric methane concentrations: Mean concentration ($\mu_s$), Standard deviation ($SD$) $\sigma_s$, Minimum and maximum values, and Range (seasonal amplitude).

This seasonal decomposition enabled the identification of temporal emission patterns across regions and provided a basis for comparative analysis. To assess seasonal variability between dairy and non-dairy regions, we conducted paired t-tests for each season. We applied a Bonferroni correction to control Type I error in multiple comparisons, setting the adjusted significance level to $\alpha = 0.0125\,(0.05\,/\,4)$. This approach aligns with established methodologies for analyzing seasonal methane emission patterns [40].

2.5.2. Seasonal Peak Identification

For each region, the weekly methane concentration time series was analyzed using a peak-finding algorithm based on local maxima detection with a minimum prominence threshold. Each identified peak was assigned to its corresponding meteorological season. To characterize the seasonal dynamics of methane emissions, the dominant peak season for each region and year was determined by analyzing the frequency of significant intra-annual peaks. A peak was classified as significant if its prominence exceeded a data-driven threshold $(T_p)$ defined as (Eq. 20):

$$T_p = \mu_P + \sigma_P \quad \ldots (20)$$

where, $\mu_P$ and $\sigma_P$ denote the mean and standard deviation of all peak prominences within the annual methane time series for the given region. The dominant season was then assigned as the meteorological season (winter, spring, summer, or fall) containing the highest number of significant peaks. This approach ensured that only statistically salient intra-annual variations were used to infer seasonal dominance, thereby enhancing resistance to noise and short-term anomalies in the time series. As a result, this methodology facilitated a robust and spatially explicit characterization of seasonal peak timing in methane concentrations, enabling inter-regional comparisons and the identification of spatiotemporal patterns in emission dynamics.

2.5.3. Seasonal Amplitude Characterization

To quantify intra-annual variation, we calculated the seasonal amplitude $(A)$ for each region as the difference between the maximum and minimum seasonal mean concentrations (Eq. 21):

$$A = \left( \max_{s \in S} \mu_S - \min_{s \in S} \mu_S \right) \quad \ldots (21)$$

where, $S$ represents the set of four meteorological seasons (winter, spring, summer, fall), and $\mu_S$ denotes the mean methane concentration for season $s$. To furthur evaluate relative seasonal dominance, we computed the pairwise seasonal concentration ratios (Eq. 22) for all pairs of seasons:

$$R_{s_1 s_2} = \frac{\mu_{s_1}}{\mu_{s_2}}, \quad \forall_{s_1 s_2} \in S, S_1 \neq S_2 \quad \ldots (22)$$

where $s_1 s_2$ are distinct seasons. These ratios provided normalized, dimensionless indicators of seasonal contrast, facilitating the identification of peak emission periods and potential drivers of seasonal methane variability. These comparative metrics are instrumental for emission source attribution and characterization of region-specific seasonal emission behaviors [41].

*2.6. Source Attribution and Emission Signature Profiling*

2.6.1. Seasonal Fingerprinting of Methane Emission Profiles

The extraction and normalization of seasonal methane patterns employed a systematic approach to identify region-specific emission signatures. Regional methane concentration data were temporally discretized into four seasonal means (winter, spring, summer, fall) for each region-year combination from 2019 to 2024. To enable interregional comparability while minimizing the influence of absolute concentrations levels, Normalized Seasonal Indices ($NSI_{r,s}$) were calculated as the ratio of each seasonal means ($\mu_{r,s}$) to the corresponding annual mean ($\mu_{r,annual}$) with values above 1.0 indicating above-average seasonal concentrations and values below 1.0 representing below-average concentrations (Eq. 23):

$$NSI_{r_1,s} = \frac{\mu_{r,s}}{\mu_{r,annual}} \quad \ldots (23)$$

where,

- $NSI_{r_1,s} = \frac{\mu_{r,s}}{\mu_{r,annual}}$ is the normalized seasonal index for region r in season s, and

- $\mu_{annual} = \frac{1}{|S|} \sum s \in S^{\mu_s}$.

These indices were then averaged across years to produce stable, four-dimensional temporal emission signatures ($NSI_{r,winter}, NSI_{r,spring}, NSI_{r,summer}, NSI_{r,fall}$) for each region. This approach standardized temporal patterns across regions with varying methane baselines, enabling the derivation of region-specific emission fingerprints through Euclidean distance based analysis.

To evaluate the distinctiveness of seasonal methane emission patterns between dairy-intensive and reference (non-dairy) regions, two quantitative approaches were employed: a multivariate hypothesis testing framework and a geometric dissimilarity metric. Hotelling's $T^2$ multivariate test was used to determine whether the multivariate seasonal profiles of dairy regions significantly differed from those of reference regions. The test statistics are defined as (Eq. 24):

$$T^2 = n \, (\overline{X}_d - \overline{X}_r)^T \, S^{-1} (\overline{X}_d - \overline{X}_r) \quad \ldots (24)$$

where, $\overline{X}_d$ and $\overline{X}_r$ are vectors of seasonal means from dairy and reference regions respectively, $S$ is the pooled covariance matrix, and $n$ is the number of samples per group. To assess statistical significance, we transformed the T² statistic into an F-distribution (Eq. 25):

$$F = \frac{n-p}{p(n-1)} T^2 \quad \ldots (25)$$

where $p = 4$ corresponds to the number of seasons, and the resulting F-statistic was evaluated at a 0.05 significance level to assess whether the differences in seasonal patterns were statistically meaningful. This technique offers robust evidence for identifying consistent seasonal patterns with subtle regional variations across climate and land use types [42,43].

In parallel, the pairwise dissimilarity between seasonal patterns was quantified using the Euclidean distance (Eq. 26) in a four-dimensional space of normalized seasonal indices (NSIs):

$$d(r_1, r_2) = \sqrt{\sum_{s \in S} (NSI_{r_1,s} - NSI_{r_2,s})^2} \quad \ldots (26)$$

where:
- $S = \{Winter, Spring, Summer, Fall\}$, and

- $d(r_1, r_2)$ denotes the dissimilarity between regions $r_1$ and $r_2$.

This distance-based approach captures relative differences in intra-annual emission structure, allowing for cross-region comparison regardless of absolute methane levels. To assess the statistical significance of the observed Euclidean distances, a permutation test was conducted with 10,000 random resamplings of region group labels [44]. This generated a null distribution of inter-group distances under the hypothesis of no difference. Empirical $p-values$ were derived by comparing the observed distances to this null distribution, providing robust inference without reliance on parametric assumptions.

2.6.2. Hierarchical Clustering of Regional Emission Signatures

To uncover spatial patterns in methane emission characteristics across regions, a hierarchical clustering analysis was conducted based on multiple standardized emission attributes [45,46]. Prior to clustering, feature normalization was applied to ensure comparability across attributes. The standardization process for each feature j across region i was defined as (Eq. 27):

$$X'_{i,j} = \frac{X_{i,j} - \mu_j}{\sigma_j} \quad \ldots (27)$$

where, $X_{i,j}$ denotes the value of feature j for region $i$, $\mu_j$ is the mean of feature $j$ across all regions, and $\sigma_j$ is the corresponding standard deviation. Agglomerative hierarchical clustering was then performed using Ward's linkage method, which minimizes the total within-cluster variance during the merging process. The dissimilarity between two clusters $C_i$ and $C_j$ was calculated as (Eq. 28):

$$D(C_i, C_j) = \frac{|C_i| \times |C_j|}{|C_i| + |C_j|} ||\mu C_i - \mu C_i||^2 \quad \ldots (28)$$

where, $|C_i|$ and $|C_j|$ represents the number of observations in cluster $C_i$ and $C_j$ respectively, and $\mu C_i$ and $\mu C_j$ are their respective centroids in the standardized feature space. To determine the optimal number of clusters, the silhouette score $s(i)$ was computed for each observation $i$ defined as (Eq. 29):

$$s(i) = \frac{b(i) - a(i)}{max\{a(i), b(i)\}} \quad \ldots (29)$$

where, $a(i)$ represents the mean intra cluster distance for point $i$, and $b(i)$ is the lowest mean distance between $i$ and any other nearest neighboring cluster. This facilitated the identification of spatially distinct clusters exhibiting similar methane emission dynamics, thereby enabling the classification of regions based on underlying dairy production practices and associated environmental drivers.

**3. Results**

3.1. Spatiotemporal Dynamics of Methane Concentrations in Canada

3.1.1. Temporal Trends in National Methane Concentrations

The national average methane concentration demonstrated a consistent upward trend over the study period (Figure 3), with an overall increase of 3.83% from 2019 (1819.79 ppb) to 2024 (1889.60 ppb), interrupted only by a notable 0.37% decrease between 2022 and 2023.

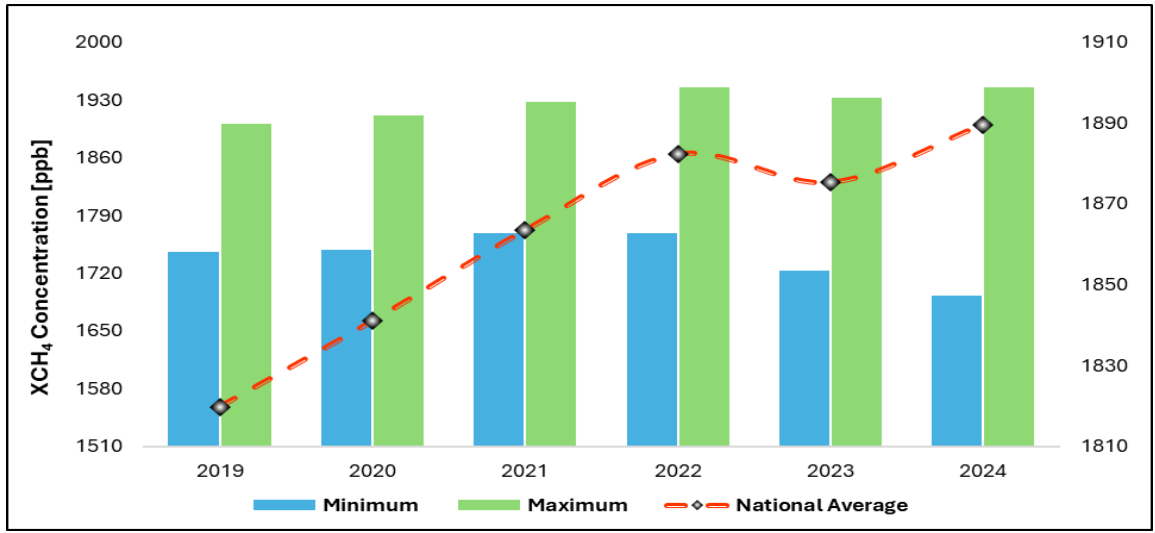

(A)

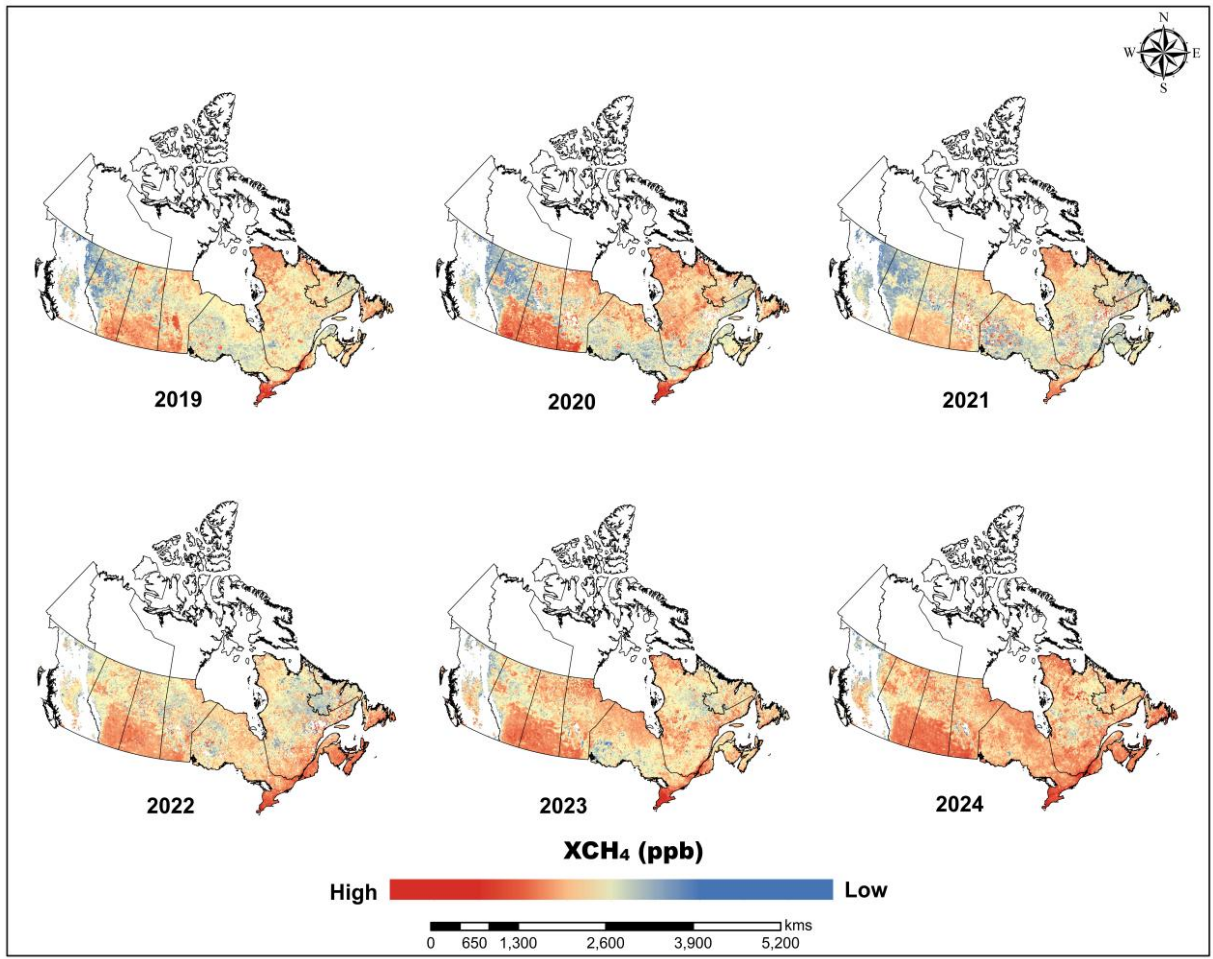

(B)

**Figure 3.** National average methane concentration showed a consistent increase from 2019 to 2024, as supported by the spatial distribution (A and B) with a notable exception between 2022 and 2023 where a slight decrease of 0.37% was observed.

This temporal pattern coincided with significant changes in the statistical distribution of methane values across provinces (Tables 2 and 3). While mean values increased steadily, we observed divergent trends in minimum and maximum values: average provincial minimum values ultimately decreased by 3.03% from 2019 to 2024 (from 1745.84 to 1692.91 ppb), while maximum values increased by 2.33% (from 1901.43 to 1945.69 ppb). Most strikingly, the average range, which represents the difference between maximum and minimum values, expanded drastically by 62.47% (from 155.59 to 252.78 ppb), indicating substantially increasing spatial heterogeneity within provinces - a trend that accelerated in the later years of the study period. The anomalous 2022-2023 decrease in national means, combined with the widening range values, suggests complex underlying dynamics potentially involving both anthropogenic and natural factors that require further investigation.

**Table 2.** National average methane concentrations and year-over-year changes, 2019-2024.

| Year | National Average (Mean) | Absolute Change | Relative Change (%) |
|---|---|---|---|
| 2019 | 1819.79 | - | - |
| 2020 | 1841.02 | +21.23 | +1.17 |
| 2021 | 1863.45 | +22.43 | +1.22 |
| 2022 | 1882.24 | +18.79 | +1.01 |
| 2023 | 1875.3 | -6.94 | -0.37 |
| 2024 | 1889.6 | +14.30 | +0.76 |
| **2019-2024** | **+69.81** | **+3.83** | **-** |

**Table 3.** Evolution of national averages of minimum, maximum, and range values, 2019-2024.

| Year | Avg. Minimum | Avg. Maximum | Avg. Range | Avg. SD |
|---|---|---|---|---|
| 2019 | 1745.84 | 1901.43 | 155.59 | 18.03 |
| 2020 | 1748.72 | 1911.1 | 162.37 | 19.48 |
| 2021 | 1768.38 | 1928.02 | 159.64 | 15.42 |
| 2022 | 1768.9 | 1945.1 | 176.2 | 13.76 |
| 2023 | 1723.39 | 1933.08 | 209.69 | 14.67 |
| 2024 | 1692.91 | 1945.69 | 252.78 | 15.67 |
| **% Change** | **-3.03%** | **+2.33%** | **+62.47%** | **-13.09%** |

3.1.2. Provincial Variations and Growth Rate Patterns

Analysis of provincial data revealed an inverse relationship between baseline methane concentrations and growth rates (Table 4 and Figure 4). Throughout the study period, Prince Edward Island consistently maintained the highest mean methane concentrations (1857.07 ppb in 2019, rising to 1913.33 ppb in 2024), while British Columbia exhibited the lowest concentrations (1789.43 ppb in 2019, increasing to 1868.70 ppb in 2024). However, British Columbia, despite having the lowest baseline concentrations, showed the highest percentage increase (4.43%) over the six-year period. Conversely, Prince Edward Island, with the highest baseline levels, demonstrated the lowest percentage increase (3.03%).

**Table 4.** Provincial methane concentrations, with absolute and percentage changes. Provinces are arranged in descending order of percentage change

| Province | Mean 2019 (ppb) | Mean 2024 (ppb) | Absolute Change (ppb) | Relative Change (%) |
|---|---|---|---|---|
| British Columbia | 1789.43 | 1868.7 | +79.27 | +4.43 |
| New Brunswick | 1818.23 | 1896.87 | +78.64 | +4.32 |
| Alberta | 1805.91 | 1883.14 | +77.22 | +4.28 |
| Ontario | 1813.64 | 1888.81 | +75.17 | +4.14 |
| Nova Scotia | 1830.6 | 1899.19 | +68.58 | +3.75 |
| Manitoba | 1816.88 | 1885.09 | +68.21 | +3.75 |
| Quebec | 1820.48 | 1887.02 | +66.54 | +3.66 |
| Saskatchewan | 1823.98 | 1889.43 | +65.46 | +3.59 |
| Newfoundland and Labrador | 1821.64 | 1884.4 | +62.76 | +3.45 |
| Prince Edward Island | 1857.07 | 1913.33 | +56.26 | +3.03 |
| National Average | 1819.79 | 1889.6 | +69.81 | +3.83 |

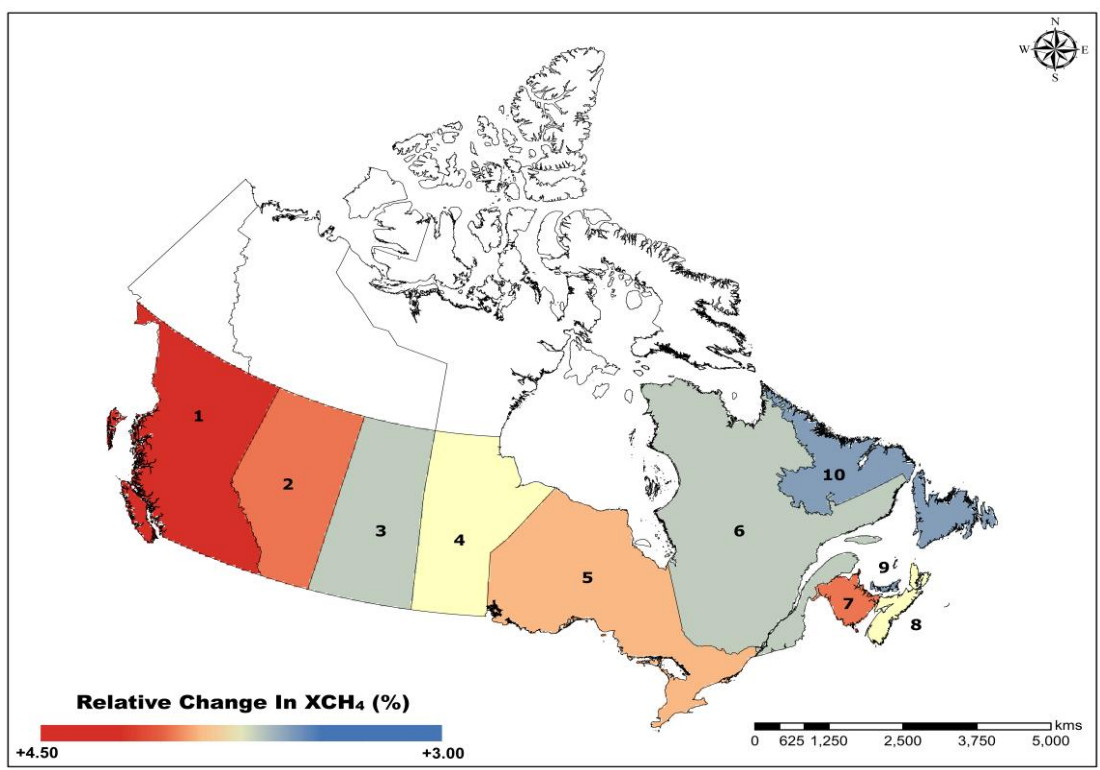

Figure 4. Spatial distribution of provincial methane concentration baseline based on relative change percentage [1 - British Columbia, 2 - Alberta, 3 - Saskatchewan, 4 - Manitoba, 5 - Ontario, 6 - Quebec, 7 - New Brunswick, 8 - Nova Scotia, 9 - Prince Edward Island, 10 - Newfoundland and Labrador].

This pattern was consistent across all provinces, with those having lower initial concentrations generally exhibiting higher growth rates. The gap between the highest and lowest provincial means narrowed over the study period, from 67.64 units in 2019 to 44.63 units in 2024, suggesting a potential convergence in provincial methane levels if current trends continue. This convergence pattern, combined with the expanding range values within provinces, indicates complex spatial dynamics in methane distribution across Canada.

*3.2. Regional Methane Concentrations Comparison and Differential Frequency*

3.2.1. Temporal Dynamics of Regional Methane Concentrations

We performed comprehensive statistical analyses of methane concentrations from dairy-intensive and non-dairy (reference) regions across Canada spanning 2019-2024 at weekly intervals. Table 5 presents the aggregate statistical parameters characterizing the complete dataset of 314 weekly observations.

Table 5. Summary statistics of atmospheric methane concentrations in diary vs. Non-dairy regions of Canada.

| Metric | Dairy Region | Non-Dairy Region | Anomaly |
|---|---|---|---|
| | (In ppb) | | |
| Mean | 1860.84 | 1843.85 | 16.99 |
| SD | 21.22 | 26.77 | 8.78 |
| Minimum | 1823.04 | 1794.87 | -6.60 |
| Maximum | 1904.32 | 1901.66 | 35.80 |

The time-domain analysis reveals a persistent elevation of methane concentrations in dairy regions compared to reference regions across the entire study period, with a mean difference of 16.99 ppb. However, this aggregate view obscures significant temporal heterogeneity in both absolute concentrations and regional differentials. To elucidate these temporal dynamics, we calculated the absolute and percentage changes between consecutive years to quantify these increases precisely, as shown in Table 6.

Table 6. Year-to-year changes and absolute changes in methane concentrations by region

| Year Transition | Dairy Avg (ppb) | Change (ppb) | % Change |
|---|---|---|---|
| 2019 | 1831.07 | - | - |
| 2019 -2020 | 1845.92 | 14.85 | 0.81 |
| 2020-2021 | 1854.72 | 8.8 | 0.48 |
| 2021-2022 | 1873.07 | 18.35 | 0.99 |
| 2022-2023 | 1877.53 | 4.46 | 0.24 |
| 2023-2024 | 1882.58 | 5.05 | 0.27 |
| **Overall Change** | - | **51.51** | **2.81** |

| Year Transition | Non-Dairy Avg (ppb) | Change (ppb) | % Change |
|---|---|---|---|
| 2019 | 1806.45 | - | - |
| 2019 -2020 | 1824.37 | 17.92 | 0.99 |
| 2020-2021 | 1835.35 | 10.98 | 0.60 |
| 2021-2022 | 1855.96 | 20.61 | 1.12 |
| 2022-2023 | 1867.47 | 11.51 | 0.62 |
| 2023-2024 | 1873.29 | 5.82 | 0.31 |
| **Overall Change** | - | **66.84** | **3.70** |

| Year Transition | Anomaly Avg (ppb) | Change (ppb) | % Change |
|---|---|---|---|
| 2019 | 24.61 | - | - |
| 2019 -2020 | 21.56 | -3.05 | -12.39 |
| 2020-2021 | 19.37 | -2.19 | -10.16 |
| 2021-2022 | 17.1 | -2.27 | -11.72 |
| 2022-2023 | 10.07 | -7.03 | -41.11 |
| 2023-2024 | 9.29 | -0.78 | -7.75 |
| **Overall Change** | - | **-15.32** | **-62.25** |

Methane levels in both dairy and non-dairy regions show a consistent upward trend from 2019 to 2024. From 2019 to 2024, dairy regions exhibited an absolute increase of 51.51 ppb in methane concentrations, representing a 2.81% relative increase. Non-dairy regions demonstrated a more pronounced growth trajectory, with an absolute increase of 66.84 ppb (3.70%). This differential growth pattern contributed to a progressive narrowing of the gap between the two region types, with the mean anomaly decreasing from 24.61 ppb in 2019 to 9.29 ppb in 2024, a 62.25% reduction over the study period.

The growth patterns were not uniform throughout the observation period. Both region types exhibited the most substantial annual increase during the 2021-2022 transition, with dairy regions increasing by 18.35 ppb (0.99%) and non-dairy regions by 20.61 ppb (1.12%). This accelerated growth period was followed by a notable deceleration in the rate of increase for dairy regions in 2022-2023 (+4.46 ppb, 0.24%) and 2023-2024 (+5.05 ppb, 0.27%).

3.2.2. Interannual Methane Variability and Regional Methane Differential Patterns

We examined the interannual variability at weekly intervals in absolute methane concentrations and the differential between dairy and non-dairy regions to identify critical transition points and pattern shifts. Figure 5 illustrates the weekly year-on-year emissions and anomaly patterns across the study period.

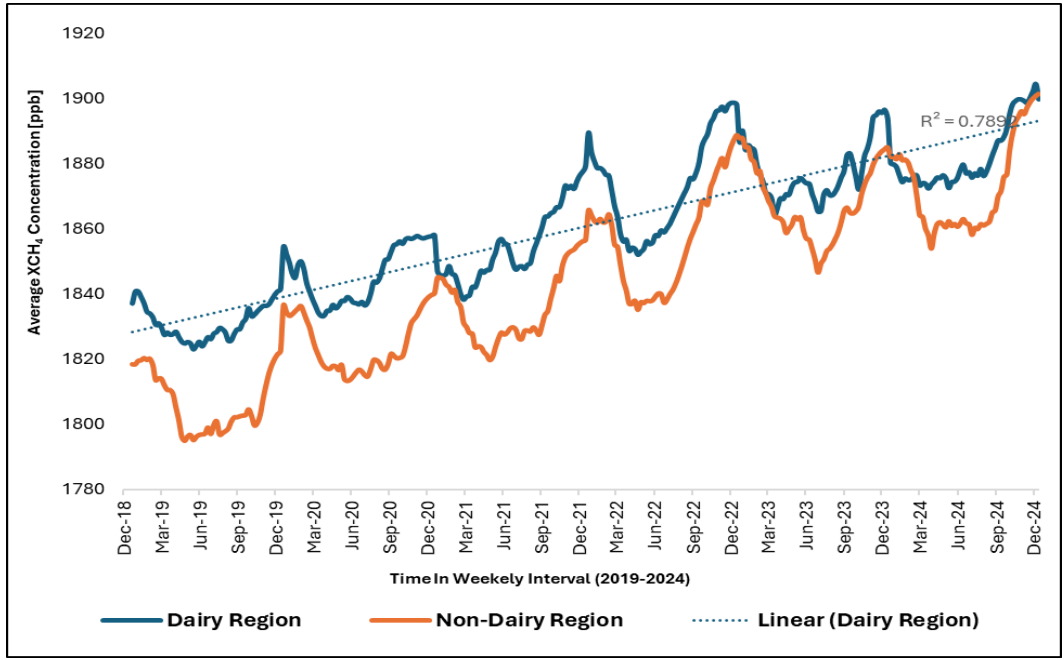

Figure 5. Weekly methane concentrations in dairy and non-dairy regions from 2019 to 2024, illustrating consistently higher methane levels in dairy regions. The correlation between methane concentrations in dairy and non-dairy regions yields an R² value of 0.78, indicating that 78% of the variability in methane levels which suggest that, despite persistent concentration differences, both region types are influenced by similar background factors.

The methane differential frequency between dairy and non-dairy regions demonstrated a consistent decreasing trend throughout the study period. However, the rate of convergence was not uniform. The 2022-2023 transition represents a critical inflection point, with the anomaly decreasing by 7.03 ppb (-41.11%), the largest single-year reduction observed in the dataset. This abrupt shift coincided with a dramatic reduction in the frequency of significant anomalies, as illustrated in Table 7. The 2022-2023 period is notably significant as it disrupts what would otherwise be a consistent increase in methane concentration. While methane levels rose overall from 2019 to 2024, the 2022-2023 period saw a 0.37% decrease in the national average (from 1882.24 ppb to 1875.3 ppb), the only year-over-year decline in the dataset. This, coupled with the accelerated convergence of methane concentrations between dairy and non-dairy regions, suggests potential common drivers that merit further investigation.

Table 7. Significant dairy-reference region methane differential frequency by year

| Year | Total Weeks | Significant Anomalies | Percentage |
|---|---|---|---|
| 2019 | 52 | 43 | 82.70 |
| 2020 | 53 | 40 | 75.50 |
| 2021 | 52 | 39 | 75.00 |
| 2022 | 52 | 24 | 46.20 |
| 2023 | 52 | 5 | 9.60 |
| 2024 | 53 | 5 | 9.40 |

To quantitatively verify the convergence between dairy and non-dairy regions, we performed statistical tests on the time series of weekly methane concentration differences. The results, presented in Table 8, provide strong statistical evidence for a significant narrowing trend in the dairy-reference methane differential throughout the study period. Both parametric (linear regression) and non-parametric (Mann-Kendall) tests confirmed a statistically significant negative trend in the methane differential. The high coefficient of determination ($R^2 = 0.88$) indicates that 88% of the variability in the dairy-reference regions convergence pattern is explained by the temporal trend. The model estimates a total reduction of 17.61 ppb in the difference between dairy and reference regions, representing a 72.32% decrease over the six-year study period. This statistically significant convergence aligns with our observation of reduced anomaly frequency and suggests a fundamental shift in the relative methane dynamics between dairy and non-dairy landscapes.

Table 8. Statistical Analysis of Dairy-Reference Methane Concentration Differential from 2019-2024

| Statistical Test | Statistic Value | Significance |
| --- | --- | --- |
| Linear Regression Slope (ppb/week) | -0.056 | $p < 0.0001$ |
| Linear Regression Slope (ppb/year) | -2.93 | $p < 0.0001$ |
| Linear Regression R² coefficient | 0.88 | - |
| Mann-Kendall Z-statistic | -20.04 | $p < 0.0001$ |
| Estimated Total Change (Absolute in ppb) | -17.61 | - |
| Estimated Total Change (Relative in %) | -72.32 | - |
| 95% Confidence Interval (Annual Slope, ppb/year) | [-3.05, -2.80] | - |

The frequency of significant anomalies, defined as methane differences exceeding a calculated threshold, decreased dramatically from 82.7% of weeks in 2019 to just 9.4% in 2024. This reduction was not gradual but occurred in distinct phases: a moderate reduction from 2019 to 2021 (82.7% to 75.0%), an accelerated decrease in 2022 (46.2%), and then a precipitous drop in 2023 (9.6%) that stabilized in 2024 (9.4%).

### 3.3. Seasonal Methane Fingerprints for Dairy Intensive Regions
#### 3.3.1. Annual Methane Concentration Patterns

Atmospheric methane concentrations exhibited distinct spatial and temporal patterns across the fourteen Canadian dairy regions during the 2019-2024 study period (Table 8). Annual mean methane concentrations ranged from 1809.58 ppb (New Brunswick Central, 2019) to 1898.02 ppb (Ontario Southern, 2024), with a multi-region average of 1860.82 ppb across the entire dataset. Table 9 illustrates the temporal evolution of methane concentrations across all major dairy provinces, revealing a consistent upward trajectory in most areas.

Table 9. Annual mean methane concentrations (ppb) by dairy region and year

| Dairy Region | 2019 | 2020 | 2021 | 2022 | 2023 | 2024 | Average |
| --- | --- | --- | --- | --- | --- | --- | --- |
| Ontario: Southern | 1853.0 | 1864.4 | 1875.1 | 1886.6 | 1891.0 | 1898.0 | 1878.0 |
| Saskatchewan: Central | 1842.5 | 1861.5 | 1870.0 | 1874.3 | 1881.8 | 1886.0 | 1869.3 |
| Alberta: Southern | 1843.5 | 1856.7 | 1871.6 | 1874.4 | 1881.0 | 1883.4 | 1868.4 |
| BC: Fraser Valley | 1840.1 | 1854.6 | 1866.8 | 1878.7 | 1883.6 | 1883.2 | 1867.8 |
| Prince Edward Island: Central | 1844.1 | 1848.9 | 1860.3 | 1881.5 | 1879.6 | 1892.2 | 1867.8 |
| Alberta: Central | 1836.7 | 1855.3 | 1866.6 | 1872.1 | 1871.6 | 1882.4 | 1864.1 |
| Manitoba: Eastern | 1832.7 | 1848.0 | 1859.5 | 1868.3 | 1879.3 | 1887.2 | 1862.5 |
| Ontario: Eastern | 1828.2 | 1841.5 | 1851.0 | 1871.2 | 1879.5 | 1891.7 | 1860.5 |
| Manitoba: Interlake | 1827.9 | 1847.7 | 1859.0 | 1868.0 | 1875.8 | 1878.8 | 1859.5 |
| Quebec: St Lawrence | 1825.3 | 1843.2 | 1846.1 | 1872.2 | 1878.5 | 1887.6 | 1858.8 |
| Quebec: East Township | 1821.7 | 1837.1 | 1844.2 | 1869.7 | 1875.5 | 1882.3 | 1855.1 |
| Nova Scotia: Valley | 1819.0 | 1829.2 | 1843.2 | 1877.8 | 1873.4 | 1884.0 | 1854.4 |
| New Brunswick Central | 1809.6 | 1826.9 | 1832.3 | 1868.2 | 1869.0 | 1878.8 | 1847.4 |
| BC: Vancouver Island | 1810.7 | 1827.9 | 1820.6 | 1859.9 | 1866.1 | 1840.5 | 1837.6 |

The spatial distribution of methane concentrations demonstrated notable regional variations, with Ontario Southern consistently maintaining the highest concentrations throughout the study period (1878.02 ppb six-year average). This was followed by Saskatchewan Central (1869.35 ppb) and Alberta Southern (1868.44 ppb), suggesting a potential correlation between intensive dairy production systems and elevated methane levels in these regions. Conversely, the Atlantic provinces, particularly New Brunswick Central (1847.45 ppb) and Nova Scotia Annapolis Valley (1854.43 ppb), exhibited consistently lower methane concentrations, which may reflect differences in dairy herd sizes, management practices, or environmental conditions affecting methane production and atmospheric dispersion.

Temporal analysis revealed that all regions except BC Vancouver Island demonstrated an overall increasing trend during the six-year period, with varying rates of change. The collective regional average increased from 1831.46 ppb in 2019 to 1882.58 ppb in 2024, representing a 2.79% increase over the study period. This progressive increase aligns with global atmospheric methane trends and underscores the contribution of dairy production to rising greenhouse gas concentrations. Interestingly, several regions exhibited a plateauing effect in 2023-2024, potentially indicating the emergence of stabilizing factors or the implementation of emission reduction strategies within the dairy sector.

3.3.2. Seasonal Amplitude and Variability Analysis

Intra-annual variability in methane concentrations, quantified through seasonal amplitude (maximum seasonal mean plus standard deviation minus minimum seasonal mean minus standard deviation), revealed distinct patterns of temporal fluctuation across the dairy regions. The six-year average seasonal amplitudes ranged from 48.13 ppb in Ontario Eastern to 83.97 ppb in BC Vancouver Island, with a multi-region mean of 61.24 ppb. This represents an average intra-annual fluctuation of approximately 3.29% relative to annual mean concentrations. Table 10 presents the spatial and temporal distribution of seasonal amplitudes across all regions and years.

| Dairy Region | 2019 | 2020 | 2021 | 2022 | 2023 | 2024 | Average |
|---|---|---|---|---|---|---|---|
| BC: Vancouver Island | 75.05 | 60.33 | 87.88 | 97.15 | 159.53 | 23.88 | 83.97 |
| Quebec: St Lawrence Valley | 45.62 | 77.50 | 62.87 | 88.25 | 51.16 | 62.67 | 64.68 |
| Quebec: Eastern Township | 26.06 | 64.47 | 59.94 | 73.11 | 63.70 | 61.65 | 58.15 |
| New Brunswick: Central | 48.53 | 39.03 | 81.48 | 97.26 | 75.44 | 56.05 | 66.30 |
| Nova Scotia Annapolis Valley | 24.49 | 56.02 | 69.55 | 97.67 | 59.15 | 61.48 | 61.39 |
| Manitoba: Interlake | 38.85 | 99.29 | 55.02 | 81.85 | 96.70 | 64.75 | 72.75 |
| Ontario: Southern | 30.47 | 61.71 | 57.90 | 56.15 | 45.57 | 66.58 | 53.06 |
| Manitoba: Eastern | 55.47 | 96.28 | 58.52 | 71.05 | 61.02 | 52.29 | 65.77 |
| Alberta: Central | 77.48 | 46.73 | 59.13 | 72.25 | 74.00 | 65.77 | 65.89 |
| Saskatchewan: Central | 53.76 | 38.13 | 39.40 | 83.80 | 43.36 | 79.54 | 56.33 |
| Prince Edward Island: Central | 47.66 | 37.83 | 80.13 | 78.83 | 33.46 | 56.23 | 55.69 |
| BC: Fraser Valley | 42.85 | 35.69 | 44.39 | 88.77 | 53.48 | 41.10 | 51.04 |
| Alberta: Southern | 49.09 | 47.74 | 51.62 | 56.68 | 58.29 | 61.58 | 54.17 |
| Ontario: Eastern | 35.03 | 34.00 | 56.79 | 79.46 | 29.66 | 53.83 | 48.13 |

Table 10. Seasonal amplitude of XCH4 (ppb) by region and year (2019-2024). The color gradient represents amplitude magnitude, with darker colors indicating higher values.

The magnitude of seasonal amplitudes exhibited substantial regional differentiation, with coastal and maritime-influenced regions generally demonstrating higher intra-annual variability. BC Vancouver Island (83.97 ppb) and Quebec St. Lawrence Valley (64.68 ppb) consistently maintained the highest seasonal amplitudes, suggesting that oceanic influence and associated meteorological patterns may amplify seasonal methane concentration cycles. Conversely, continental regions such as Ontario Eastern (48.13 ppb) and Alberta Southern (54.17 ppb) exhibited more moderate seasonal variations, indicating potentially more stable emission patterns throughout the year or different atmospheric dispersion characteristics. Temporal analysis of seasonal amplitudes revealed notable year-to-year fluctuations without consistent trends across the study period. The collective regional average seasonal amplitude increased from 46.46 ppb in 2019 to 57.67 ppb in 2024, representing a 24.13% increase in intra-annual variability. However, this overall increase masks substantial inter-annual variations, with 2022 exhibiting the highest mean regional amplitude (80.16 ppb) and 2019 the lowest (46.46 ppb). Of particular note was BC Vancouver Island in 2023, which exhibited an exceptional seasonal amplitude of 159.53 ppb, representing an 8.56% fluctuation relative to its annual mean - the largest observed in the entire dataset.

The absence of consistent temporal trends in seasonal amplitude suggests that factors governing intra-annual variability may be influenced more by year-specific meteorological conditions than by systematic changes in dairy management practices. The synchronized peak in amplitude across ten regions during 2022 indicates a potential continental-scale meteorological influence affecting methane dispersion during that year. However, the general increase in average amplitude from 2019 to 2024 warrants further investigation as it may indicate increasing seasonal extremes in emission patterns, potentially related to climate change effects on dairy production systems.

3.3.3. Phase Shift Analysis and Timing of Peak Emissions

The seasonal distribution of methane concentrations revealed consistent temporal patterns in emission peaks across the study regions. Analysis of seasonal maxima identified fall (September-November) and winter (December-February) as the predominant periods of peak methane emissions, with specific regional preferences evident in the data. Figure 6 illustrates the seasonal distribution of peak emissions by region and year, highlighting the temporal consistency of these patterns.

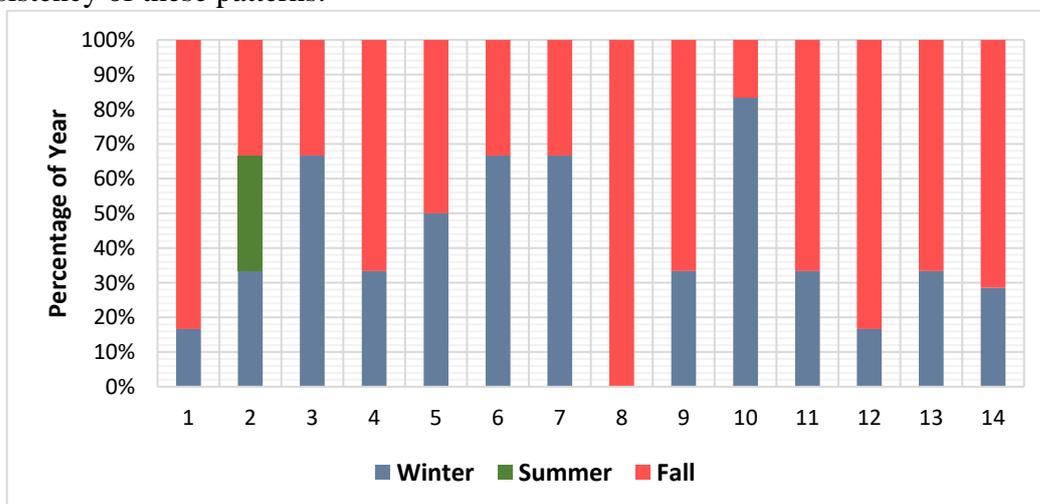

Figure 6. Seasonal distribution of peak methane emissions by region (2019-2024). The stacked bar chart shows the distribution of peak seasons for each region. Each bar represents a dairy region (1 - BC: Fraser Valley, 2 - BC: Vancouver Island, 3 - Alberta: Central, 4 - Alberta: Southern, 5 - Saskatchewan: Central, 6 - Manitoba: Interlake, 7 - Manitoba: Eastern, 8 - Ontario: Southern, 9 - Ontario: Eastern, 10 - Quebec: St Lawrence Valley, 11 - Quebec: Eastern Township, 12 - New Brunswick: Central, 13 - Nova Scotia: Annapolis Valley, 14 - Prince Edward Island: Central) with different colors indicating the proportion of years where winter, spring, summer, or fall was the peak season.

Fall emerged as the dominant peak season in eight of the fourteen regions (57.1%), with a particularly strong fall dominance observed in Ontario Southern (100% of years), BC Fraser

Valley (83.3% of years), and Prince Edward Island Central (83.3% of years). The consistent fall peak in these regions suggests a potential link to specific agricultural activities occurring during this period, such as manure application before winter freeze, changes in feeding regimes as cattle transition from grazing to confined housing or reduced atmospheric mixing heights limiting methane dispersion. Winter dominated as the peak emission season in six regions (42.9%), including Quebec St. Lawrence Valley (83.3% of years), Manitoba Interlake (66.7% of years), Manitoba Eastern (66.7% of years), and Alberta Central (66.7% of years). These winter-dominant regions share commonalities in continental climate conditions with extended cold periods, suggesting that winter housing of cattle and associated manure management practices may significantly influence methane emission patterns. The winter peak may also reflect reduced methane oxidation in frozen soils and limited atmospheric dispersion under winter inversion conditions.

Notably, spring (March-May) never emerged as the peak season in any region throughout the study period, consistently showing the lowest methane concentrations across all areas. This universal spring minimum suggests fundamental processes limiting methane emissions during this period, potentially including increased soil methane oxidation as temperatures rise, different manure management practices during the planting season, or increased atmospheric mixing and dispersion during spring. The temporal consistency of seasonal patterns within regions, with limited year-to-year variation in peak timing, indicates that the seasonal dynamics of methane emissions are governed by stable regional factors rather than stochastic influences. This temporal consistency strengthens the potential utility of seasonal patterns as diagnostic "fingerprints" for dairy emission sources.

3.3.4. Regional Fingerprint Derivation and Analysis

To establish a subtly different regional emission signatures, we calculated normalized seasonal indices as the ratio of seasonal mean concentrations to annual means for each region (Table11).

Table 11. Normalized Seasonal Indices (Seasonal Mean/Annual Mean) by region:

| Dairy Region | Winter/Annual | Spring/Annual | Summer/Annual | Fall/Annual |
|---|---|---|---|---|
| BC: Fraser Valley | 0.9988 | 0.9987 | 0.9978 | 1.0047 |
| BC: Vancouver Island | 1.0035 | 0.9967 | 0.9996 | 1.0007 |
| Alberta: Central | 1.0047 | 0.9963 | 0.995 | 1.0044 |
| Alberta: Southern | 1.0029 | 0.9954 | 0.9962 | 1.006 |
| Saskatchewan: Central | 1.0046 | 0.995 | 0.9958 | 1.0052 |
| Manitoba: Interlake | 1.0046 | 0.9938 | 0.9955 | 1.0039 |
| Manitoba: Eastern | 1.0074 | 0.9939 | 0.9956 | 1.0045 |
| Ontario: Southern | 1.0064 | 0.9967 | 0.9978 | 1.0061 |
| Ontario: Eastern | 0.9995 | 0.996 | 0.9975 | 1.0048 |
| Quebec: St Lawrence Valley | 1.002 | 0.9934 | 0.9963 | 1.004 |
| Quebec: Eastern Township | 1.0068 | 0.9941 | 0.9965 | 1.0056 |
| New Brunswick: Central | 1.0043 | 0.9937 | 0.997 | 1.0054 |

| Nova Scotia Annapolis Valley | 1.0045 | 0.9942 | 0.9975 | 1.0052 |
| Prince Edward Island: Central | 1.0036 | 0.995 | 0.997 | 1.0065 |

These normalized indices facilitate direct comparison of seasonal patterns while controlling differences in absolute concentration levels. Figure 7 presents the normalized seasonal fingerprints for four regionally representative dairy regions, revealing subtle but diagnostically significant variations in emission patterns in major dairy regions. The normalized seasonal indices revealed a consistent fundamental pattern across all regions, characterized by below-average concentrations in spring and summer, contrasted with above-average concentrations in fall and winter. This universal pattern was modulated by region-specific variations that constituted the subtly different regional fingerprints of each area. All regions exhibited spring minima, with spring-to-annual ratios ranging from 0.9934 (Quebec St. Lawrence Valley) to 0.9987 (BC Fraser Valley), representing a 0.53% spread across regions. Similarly, summer concentrations were consistently below annual averages, with summer-to-annual ratios ranging from 0.9950 (Alberta Central) to 0.9996 (BC Vancouver Island).

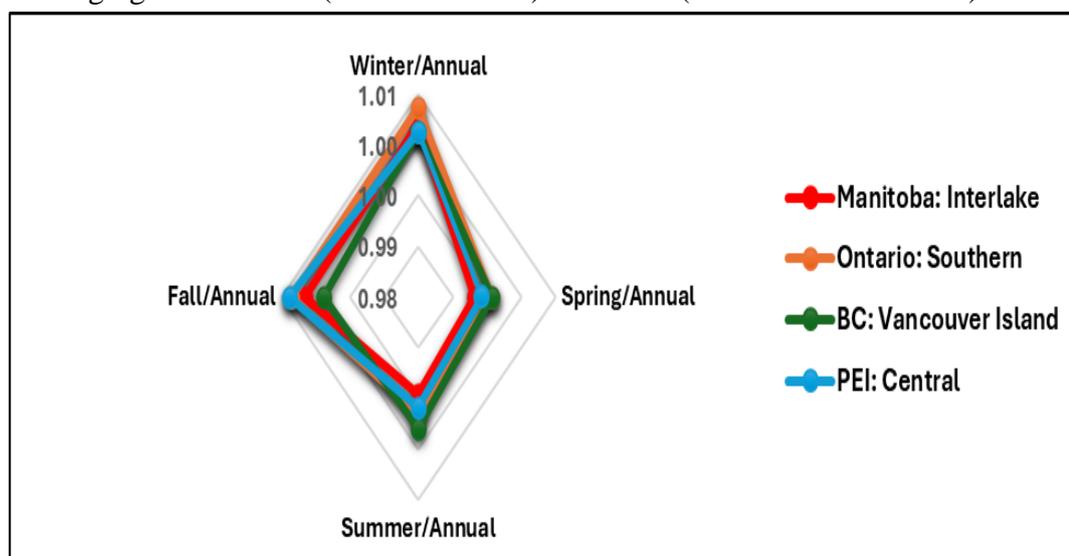

Figure 7. Regional methane emission fingerprints based on normalized seasonal indices for representative dairy emission regions. Manitoba: Eastern (highest winter peak), Ontario: Eastern (winter depression), BC: Vancouver Island (most balanced pattern), and Prince Edward Island: Central (strongest fall peak). The radial values range from 0.99 to 1.01, showing seasonal deviations from annual mean concentrations.

The elevated methane concentrations during winter and fall exhibited greater regional differentiation, serving as key diagnostic components of the seasonal regional signatures. Winter-to-annual concentration ratios ranged from 0.9988 (British Columbia, Fraser Valley) to 1.0074 (Manitoba, Eastern), corresponding to a spread of 0.86%. Similarly, fall-to-annual ratios ranged from 1.0007 (British Columbia, Vancouver Island) to 1.0065 (Prince Edward Island, Central), yielding a spread of 0.58%. Although these seasonal variations were modest (0.5–1%), they were statistically significant in several pairwise comparisons ($p < 0.05$), indicating subtle but meaningful regional differences. Overall, the high degree of consistency in seasonal indices across dairy regions suggests common underlying mechanisms influencing methane dynamics within Canada's dairy sector. Prairie regions (Manitoba, Saskatchewan, and Alberta) exhibited strong winter maxima coupled with pronounced spring minima, a pattern consistent with continental climate influences on both emission processes and atmospheric dynamics. Atlantic regions (Prince Edward Island, Nova Scotia, and New Brunswick) demonstrated a consistent pattern of fall dominance with moderate winter maxima. Ontario and Quebec regions showed mixed patterns, potentially reflecting their transitional position between continental and maritime influences.

The most distinctive regional fingerprints included Manitoba Eastern, which exhibited the highest winter maximum (1.0074) coupled with a substantial spring minimum (0.9939), representing a 1.35% seasonal swing. Ontario Eastern demonstrated a unique pattern with a slight winter depression (0.9995) contrasted with a strong fall maximum (1.0048). Ontario Southern showed high values in both winter (1.0064) and fall (1.0061), the second highest values in both seasons. BC Vancouver Island displayed the most balanced seasonal pattern with the smallest amplitude between seasonal indices (0.0068, calculated as the difference between maximum and minimum normalized indices), consistent with its maritime climate moderating seasonal extremes. Quantitative similarity analysis using Euclidean distance metrics confirmed the high level of consistency in seasonal patterns across all regions, with all paired-region distances below 0.01 (Figure 6). The closest similarities were observed between geographically proximate regions, including Quebec Eastern Townships and New Brunswick Central (distance: 0.0007), and Manitoba Interlake and Manitoba Eastern (distance: 0.0012). Despite these variations, the overall pattern of spring minima and fall-winter maxima remains remarkably consistent across all regions, suggesting similar underlying drivers of seasonal methane dynamics throughout Canada's dairy sector. This pattern of distance-based clustering provides quantitative evidence that while seasonal patterns follow a common fundamental structure, subtle regional "fingerprints" do emerge that reflect local climate and potentially management conditions.

### 4. Discussion

In this study, we analyzed six years of satellite methane observations over Canadian dairy farming regions and selected non-dairy farming regions, revealing notable patterns. We found that atmospheric methane concentrations increased nationwide from 2019 to 2024, but with significant regional variations. Most notably, we observed a substantial narrowing (62.25%) in the methane concentration difference between dairy and non-dairy regions, suggesting complex emission dynamics. We also identified consistent seasonal patterns across dairy regions, with fall-winter peaks and spring minima, creating subtle regional "fingerprints" that could support source attribution efforts. Our analysis of Sentinel-5P TROPOMI data revealed significant methane concentration patterns across Canadian dairy regions. The national upward trend of 3.83% aligns with global increases documented by [2], while [47] call for finer spatial resolution in methane monitoring. The brief 0.37% decrease between 2022-2023 suggests interannual variability influenced by climatic anomalies [15]. The expansion in concentration range (62.47%) indicates increasing spatial heterogeneity despite converging provincial means, contradicting [48] suggestion of uniform growth patterns in northern latitudes. We observed an inverse relationship between baseline concentrations and growth rates at the provincial level.

British Columbia (1789.43 ppb in 2019) showed higher increases (4.43%) than Prince Edward Island (1857.07 ppb, 3.03% increase). This pattern aligns with [49] findings but contrasts with [15], suggesting unique regional dynamics in Canadian agricultural zones. The varying temporal stability across provinces supports [50] hypothesis that methane dynamics reflect complex environmental interactions, consistent with [51] documentation of stronger seasonal patterns in coastal regions. Our quasi-experimental comparison revealed a surprising convergence between dairy and non-dairy regions, with methane concentration anomaly narrowing from 24.61 ppb (2019) to 9.29 ppb (2024), a 62.25% reduction. This observed trend contrasts with earlier model-based projections, such as those discussed by [52], which suggested persistently higher methane levels in dairy-dense regions due to underestimated agricultural emissions. Statistical analysis of the weekly methane differential time series provided a robust quantitative support for this convergence pattern, with a highly significant negative trend (-2.93 ppb per year, $p<0.0001$) and strong temporal dependence ($R^2 = 0.88$). The estimated total change in the differential (-17.61 ppb) represented a 72.32% reduction over the study period, highlighting a notable magnitude of convergence between dairy and non-diary regions as often seen in methane emission trend dynamics [53]. Importantly, this convergence does not imply a decline in absolute methane concentrations within dairy regions, but rather reflects differential growth rates, with methane levels in non-dairy regions increasing slightly faster (3.70% from 2019 to 2024) compared to dairy regions (2.81%). This differential growth contributed to a marked reduction in the baseline concentration gap between dairy and non-dairy areas. Notably, absolute methane concentrations in dairy regions continued to rise by approximately 2.8% over the study period, although at a slower rate than in non-dairy regions. The period from 2022 to 2023 emerged as a critical transition point, characterized by a substantial 41.11% decrease in the frequency of methane anomalies. While [9] noted regional gradient shifts due to meteorological anomalies, our observed convergence may reflect broader factors including increased high-latitude wetland emissions [54] or meteorological influences [55]. Despite the convergence trend, the observed mean methane enhancement over dairy regions (16.99 ppb) aligns with inventory-based expectations. [6] estimates 23 gigagram (Gg) $CH_4$ emissions from the dairy sector in 2021, primarily from enteric fermentation and manure management. When spatially distributed across the 2° × 2° grid dairy-intensive areas, this emission magnitude stand consistent with the satellite-derived enhancements observed. This consistency can be understood through atmospheric mixing and dispersion principles, where point-source emissions are diluted within the atmospheric column measured by Sentinel-5P TROPOMI. Seasonal analysis revealed distinctive methane "fingerprints" with fall and winter consistently showing peak concentrations. [56] documented similar patterns due to manure storage practices and reduced atmospheric mixing. [57] and [58] observed elevated enteric methane from dairy cattle during winter due to feed composition and housing changes. The universal spring minimum suggests fundamental seasonal constraints on emissions, consistent with [59] but contradicting [14] observations of lagged environmental responses in tropical systems. The normalized seasonal indices showed significant regional variations supporting [56] conclusions about regional climate factors, addressing [60] critique of current attribution approaches. Regionally, Ontario Southern maintained consistently high concentrations (1878.02 ppb six-year average), followed by Saskatchewan Central (1869.35 ppb) and Alberta Southern (1868.44 ppb). This aligns [56] identification of highest dairy cattle densities and supports [58] spatial association between livestock density and methane enhancements however, [61] argued that wetland emissions may dominate these regional budgets. We found greater seasonal variability in coastal regions (BC Vancouver Island: 68.12 ppb) than continental areas (Ontario Eastern: 39.23 ppb), supporting [60] observations but contradicting [15] continental amplification hypothesis. Our methane emission data reflects multiple processes including emissions, transport, and transformations. However, [12,13] cautioned against simple attribution without considering meteorological effects. The dairy-non-dairy convergence may reflect changes in natural methane sources, supported by [62] work of increasing environmental emissions, though [9] found limited evidence for substantial natural emission increases in agricultural regions. BC Vancouver Island's exceptional 2023 seasonal amplitude (146.66 ppb) exemplifies how meteorological anomalies influence methane dynamics [55], though [63] argued such anomalies primarily reflect emission process changes. The seasonal fingerprints likely reflect agricultural practices and environmental factors. [64] documented strong seasonal patterns in Canadian dairy emissions linked to management practices and environmental conditions. [65] found seasonal variations driven by feeding regimes and temperature responses. Regional differences in seasonal patterns align with [66] findings, though [13] argued atmospheric transport patterns may drive observed variations. The spring minimum likely reflects soil methane oxidation and different manure management [7], or enhanced atmospheric mixing [59]. The convergence between regions requires careful interpretation within Canadian climate policy context as the 2022-2023 acceleration coincides with enhanced climate policies [6] which also aligns with the reported emission reductions in Canada [67]. However, [68,69] found limited evidence for rapid policy-driven reductions, suggesting alternative explanations including increased wetland emissions [54] or atmospheric transport changes [55]. High interannual variability [48] complicates attribution while the consistent regional fingerprints support [5] conclusions about standardized management practices, though [14] suggested apparent consistency may reflect atmospheric processes. The subtle but significant regional pattern variations validates [70] work on seasonal characteristic information, though [71] cautioned against over-interpretation. The geographical clustering reinforces [72] emphasis on regional mitigation approaches, aligning with Canadian modeling studies [73,74]. These regional signatures provide a foundation for targeted strategies addressing [1] recommendations, supported by [15] emphasis on temporally targeted mitigation focusing on peak emission periods.

## 5. Conclusions

This study presents the first comprehensive satellite-based assessment of methane dynamics in Canadian dairy regions, revealing complex spatiotemporal patterns that challenge simplistic assumptions about agricultural emissions. Our findings demonstrate a consistent national increase in atmospheric methane concentrations from 2019 to 2024, with significant regional and seasonal variations providing insights into underlying drivers of methane flux. The most notable finding is the 62.25% reduction in the methane concentration anomaly between dairy and non-dairy regions, which challenges traditional attribution models and indicates significant changes in emission dynamics. Our statistical analysis provides robust quantitative evidence for this convergence, demonstrating a highly significant negative trend in the methane differential time series (-2.93 ppb/year, p<0.0001) that explains 88% of the observed variance. This narrowing gap may be attributed to increases in baseline methane levels (e.g., due to wetland activity or climate-driven factors), rather than a definitive decline in emissions from dairy sources. The identification of 2022-2023 as a critical transition period offers a temporal focus for further investigation into the underlying drivers of this abrupt change. The consistent seasonal methane patterns observed across dairy regions, with their subtle regional variations, provide valuable insights for understanding dairy emission dynamics and temporal targeting of mitigation strategies. While the small differences between regions may have limited utility for source attribution within the dairy sector, the characteristic seasonal pattern of dairy emissions (fall-winter maximum, spring minimum) could potentially be distinguished from other major methane sources with different seasonal dynamics in future comparative studies. The convergence between dairy and non-dairy regions suggests either that mitigation efforts may be yielding positive results or that more complex atmospheric processes are at play. This emphasizes the importance of comprehensive monitoring systems that integrate satellite observations with ground-based measurements and process-based models to accurately attribute methane sources and evaluate mitigation efficacies. The regional and seasonal specificity of our results underscores the need for targeted interventions that address peak emission periods and account for local environmental conditions. Future research should focus on validating these satellite-derived patterns through ground-based measurements and exploring the mechanistic drivers behind the observed convergence trend. The methodology developed in this study offers a powerful framework for monitoring agricultural greenhouse gas emissions at regional scales, with potential applications beyond the dairy sector to other agricultural and natural methane sources.

**Author Contributions:** Conceptualization, S.N. and P.J.P..; methodology, P.J.P.; validation, P.J.P., S.N. and M.G.; formal analysis, P.J.P; investigation, P.J.P.; resources, S.N.; writing—original draft preparation, P.J.P.; writing—review and editing, P.J.P.; K.P.; M.G.; S.N.; visualization, S.N.; supervision, S.N.; project administration, S.N.; funding acquisition, S.N. All authors have read and agreed to the published version of the manuscript.
**Funding:** This work is kindly sponsored by the Natural Sciences and Engineering Research Council of Canada (RGPIN 2024-04450), Net Zero Atlantic Canada Agency (300700018), Mitacs Canada (IT36514), and the Department of New Brunswick Agriculture, Aquaculture and Fisheries (NB2425-0025).
**Data Availability Statement:** The data is available from the corresponding author upon reasonable request.
**Conflicts of Interest:** The authors declare no conflicts of interest.

## References

1. IPCC Climate Change 2022: Mitigation of Climate Change. Contribution of Working Group III to the Sixth Assessment Report of the Intergovernmental Panel on Climate Change; Cambridge University Press: Cambridge, UK and New York, NY, USA, 2023; ISBN 978-92-9169-160-9.[Google Scholar]
2. Nisbet, E.G.; Manning, M.R.; Dlugokencky, E.J.; Fisher, R.E.; Lowry, D.; Michel, S.E.; Myhre, C.L.; Platt, S.M.; Allen, G.; Bousquet, P.; Brownlow, R. Very Strong Atmospheric Methane Growth in the 4

Years 2014–2017: Implications for the Paris Agreement. Global Biogeochemical Cycles **2019**, 33, 318–342. [Google Scholar]

3. Friedlingstein, P.; Jones, M.W.; O'Sullivan, M.; Andrew, R.M.; Bakker, D.C.E.; Hauck, J.; Le Quéré, C.; Peters, G.P.; Peters, W.; Pongratz, J.; et al. Global Carbon Budget 2021. Earth Syst. Sci. Data **2022**, 14, 1917–2005, doi:10.5194/essd-14-1917-2022. [Google Scholar]

4. Crippa, M.; Solazzo, E.; Guizzardi, D.; Monforti-Ferrario, F.; Tubiello, F.N.; Leip, A. Food Systems Are Responsible for a Third of Global Anthropogenic GHG Emissions. Nat Food **2021**, 2, 198–209. [Google Scholar]

5. Huang, D.; Guo, H. Diurnal and Seasonal Variations of Greenhouse Gas Emissions from a Naturally Ventilated Dairy Barn in a Cold Region. Atmospheric Environment **2018**, 172, 74–82, d. [Google Scholar]

6. Environment and Climate Change Canada National Inventory Report 1990–2021: Greenhouse Gas Sources and Sinks in Canada. Canada's Submission to the United Nations Framework Convention on Climate Change; Environment and Climate Change Canada: Gatineau, QC, Canada, 2023;

7. Statistics Canada Canada's 2021 Census of Agriculture: A Story About the Transformation of the Agriculture Industry and Adaptiveness of Canadian Farmers; 2022; pp. 1–12;.

8. Schissel, C.; Allen, D.; Dieter, H. Methods for Spatial Extrapolation of Methane Measurements in Constructing Regional Estimates from Sample Populations. Environ. Sci. Technol. **2024**, 58, 2739–2749. [Google Scholar]

9. Wolf, J.; Asrar, G.R.; West, T.O. Revised Methane Emissions Factors and Spatially Distributed Annual Carbon Fluxes for Global Livestock. Carbon Balance Manage **2017**, 12, 16. [Google Scholar]

10. Soranno, P.A.; Wagner, T.; Collins, S.M.; Lapierre, J.; Lottig, N.R.; Oliver, S.K. Spatial and Temporal Variation of Ecosystem Properties at Macroscales. Ecology Letters **2019**, 22, 1587–1598. [Google Scholar]

11. Chan, E.; Worthy, D.E.J.; Chan, D.; Ishizawa, M.; Moran, M.D.; Delcloo, A.; Vogel, F. Eight-Year Estimates of Methane Emissions from Oil and Gas Operations in Western Canada Are Nearly Twice Those Reported in Inventories. Environ. Sci. Technol. **2020**, 54, 14899–14909. [Google Scholar]

12. Jacob, D.J.; Varon, D.J.; Cusworth, D.H.; Dennison, P.E.; Frankenberg, C.; Gautam, R.; Guanter, L.; Kelley, J.; McKeever, J.; Ott, L.E.; Poulter, B. Quantifying Methane Emissions from the Global Scale down to Point Sources Using Satellite Observations of Atmospheric Methane. Atmos. Chem. Phys. **2022**, 22, 9617–9646. [Google Scholar]

13. Pandey, S.; Gautam, R.; Houweling, S.; Van Der Gon, H.D.; Sadavarte, P.; Borsdorff, T.; Hasekamp, O.; Landgraf, J.; Tol, P.; Van Kempen, T.; Hoogeveen, R. Satellite Observations Reveal Extreme Methane Leakage from a Natural Gas Well Blowout. Proc. Natl. Acad. Sci. U.S.A. **2019**, 116, 26376–26381. [Google Scholar]

14. Bloom, A.A.; Bowman, K.W.; Liu, J.; Konings, A.G.; Worden, J.R.; Parazoo, N.C.; Meyer, V.; Reager, J.T.; Worden, H.M.; Jiang, Z.; Quetin, G.R. Lagged Effects Regulate the Inter-Annual Variability of the Tropical Carbon Balance. Biogeosciences **2020**, 17, 6393–6422. [Google Scholar]

15. Zhang, Y.; Jacob, D.J.; Lu, X.; Maasakkers, J.D.; Scarpelli, T.R.; Sheng, J.-X.; Shen, L.; Qu, Z.; Sulprizio, M.P.; Chang, J.; Bloom, A.A. Attribution of the Accelerating Increase in Atmospheric Methane during 2010–2018 by Inverse Analysis of GOSAT Observations. Atmos. Chem. Phys. **2021**, 21, 3643–3666. [Google Scholar]

16. Maasakkers, J.D.; Jacob, D.J.; Sulprizio, M.P.; Scarpelli, T.R.; Nesser, H.; Sheng, J.; Zhang, Y.; Lu, X.; Bloom, A.A.; Bowman, K.W.; Worden, J.R. 2010–2015 North American Methane Emissions, Sectoral Contributions, and Trends: A High-Resolution Inversion of GOSAT Observations of Atmospheric Methane. Atmos. Chem. Phys. **2021**, 21, 4339–4356. [Google Scholar]
17. East, J.D.; Jacob, D.J.; Balasus, N.; Bloom, A.A.; Bruhwiler, L.; Chen, Z.; Kaplan, J.O.; Mickley, L.J.; Mooring, T.A.; Penn, E.; Poulter, B. Interpreting the Seasonality of Atmospheric Methane. Geophysical Research Letters **2024**, 51, e2024GL108494. [Google Scholar]
18. Canadian Dairy Commission Fact Sheet - Supply Management Available online: https://www.cdc-ccl.ca/en/node/890.
19. Hasekamp, O.; Lorente, A.; Hu, H.; Butz, A.; de Brugh, J.; Landgraf, J. Algorithm Theoretical Baseline Document for Sentinel-5 Precursor Methane Retrieval; Netherlands Institute for Space Research: The Netherlands, 2021; p. 67. [Google Scholar]
20. Häme, T.; Sirro, L.; Kilpi, J.; Seitsonen, L.; Andersson, K.; Melkas, T. A Hierarchical Clustering Method for Land Cover Change Detection and Identification. Remote Sensing **2020**, 12, 1751. [Google Scholar]
21. Zou, Y.; Greenberg, J.A. A Spatialized Classification Approach for Land Cover Mapping Using Hyperspatial Imagery. Remote Sensing of Environment **2019**, 232, 111248. [Google Scholar]
22. Agriculture and Agri-Food Canada Number of Farms with Dairy Cows and Dairy Heifers Available online: https://agriculture.canada.ca/en/sector/animal-industry/canadian-dairy-information-centre/dairy-statistics-and-market-information/farm-statistics/number-farms-dairy-cows-and-dairy-heifers.
23. IPCC 2019 Refinement to the 2006 IPCC Guidelines for National Greenhouse Gas Inventories; Intergovernmental Panel on Climate Change: Switzerland;
24. Maasakkers, J.D.; Jacob, D.J.; Sulprizio, M.P.; Turner, A.J.; Weitz, M.; Wirth, T.; Hight, C.; DeFigueiredo, M.; Desai, M.; Schmeltz, R.; et al. Gridded National Inventory of U.S. Methane Emissions. Environ. Sci. Technol. **2016**, 50, 13123–13133. [Google Scholar]
25. Moon, J.; Shim, C.; Seo, J.; Han, J. Evaluation of Korean Methane Emission Sources with Satellite Retrievals by Spatial Correlation Analysis. Environ Monit Assess **2024**, 196, 296. [Google Scholar]
26. Ester, M.; Kriegel, H.P.; Sander, J.; Xu, X. A Density-Based Algorithm for Discovering Clusters in Large Spatial Databases with Noise. In Proceedings of the Proceedings of the 2nd International Conference on Knowledge Discovery and Data Mining (KDD); Portland, OR, USA, August 1996; Vol. 96, pp. 226–231. [Google Scholar]
27. Chen, B.Y.; Luo, Y.-B.; Zhang, Y.; Jia, T.; Chen, H.-P.; Gong, J.; Li, Q. Efficient and Scalable DBSCAN Framework for Clustering Continuous Trajectories in Road Networks. International Journal of Geographical Information Science **2023**, 37, 1693–1727. [Google Scholar]
28. Park, J.Y.; Ryu, D.J.; Nam, K.W.; Jang, I.; Jang, M.; Lee, Y. DeepDBSCAN: Deep Density-Based Clustering for Geo-Tagged Photos. IJGI **2021**, 10, 548. [Google Scholar]
29. Boeing, G. Clustering to Reduce Spatial Data Set Size. SSRN Journal **2018**. [Google Scholar]
30. Cusworth, D.H.; Jacob, D.J.; Sheng, J.-X.; Benmergui, J.; Turner, A.J.; Brandman, J.; White, L.; Randles, C.A. Detecting High-Emitting Methane Sources in Oil/Gas Fields Using Satellite Observations. Atmos. Chem. Phys. **2018**, 18, 16885–16896. [Google Scholar]

31. Varon, D.J.; Jervis, D.; McKeever, J.; Spence, I.; Gains, D.; Jacob, D.J. High-Frequency Monitoring of Anomalous Methane Point Sources with Multispectral Sentinel-2 Satellite Observations. Atmos. Meas. Tech. **2021**, 14, 2771–2785, doi:10.5194/amt-14-2771-2021. [Google Scholar]
32. Jongaramrungruang, S.; Matheou, G.; Thorpe, A.K.; Zeng, Z.-C.; Frankenberg, C. Remote Sensing of Methane Plumes: Instrument Tradeoff Analysis for Detecting and Quantifying Local Sources at Global Scale. Atmos. Meas. Tech. **2021**, 14, 7999–8017. [Google Scholar]
33. Welch, G.; Bishop, G. An Introduction to the Kalman Filter; University of North Carolina at Chapel Hill, 1995. [Google Scholar]
34. Grewal, M.S.; Andrews, A.P. Kalman Filtering: Theory and Practice with MATLAB; 4th ed.; John Wiley & Sons: Hoboken, NJ, USA, 2014. [Google Scholar]
35. Steiner, M.; Cantarello, L.; Henne, S.; Brunner, D. Flow-Dependent Observation Errors for Greenhouse Gas Inversions in an Ensemble Kalman Smoother. Atmos. Chem. Phys. **2024**, 24, 12447–12463, doi:10.5194/acp-24-12447-2024. [Google Scholar]
36. Dlugokencky, E.; Crotwell, A.; Masarie, K.; White, J.; Lang, P.; Crotwell, M. NOAA Measurements of Long-Lived Greenhouse Gases. Asia-Pacific GAW Greenhouse Gases **2013**, 6, 6–9. [Google Scholar]
37. Hamed, K.H.; Ramachandra Rao, A. A Modified Mann-Kendall Trend Test for Autocorrelated Data. Journal of Hydrology **1998**, 204, 182–196. [Google Scholar]
38. Barro, R.J. Convergence. Journal of Political Economy **1992**, 100, 223–251. [Google Scholar]
39. NCEI Meteorological versus Astronomical Seasons Available online: https://www.ncei.noaa.gov/news/meteorological-versus-astronomical-seasons (accessed on 1 March 2025).
40. Trisna, B.A.; Park, S.; Lee, J. Significant Impact of the Covid-19 Pandemic on Methane Emissions Evaluated by Comprehensive Statistical Analysis of Satellite Data. Sci Rep **2024**, 14, 22475. [Google Scholar]
41. Karoff, C.; Vara-Vela, A.L. Data Driven Analysis of Atmospheric Methane Concentrations as Function of Geographic, Land Cover Type and Season. Front. Earth Sci. **2023**, 11, 1119977. [Google Scholar]
42. Roth, F.; Sun, X.; Geibel, M.C.; Prytherch, J.; Brüchert, V.; Bonaglia, S.; Broman, E.; Nascimento, F.; Norkko, A.; Humborg, C. High Spatiotemporal Variability of Methane Concentrations Challenges Estimates of Emissions across Vegetated Coastal Ecosystems. Global Change Biology **2022**, 28, 4308–4322. [Google Scholar]
43. DelSole, T.; Tippett, M.K. Comparing Climate Time Series – Part 2: A Multivariate Test. Adv. Stat. Clim. Meteorol. Oceanogr. **2021**, 7, 73–85. [Google Scholar]
44. Anderson, M.J. Distance-Based Tests for Homogeneity of Multivariate Dispersions. Biometrics **2006**, 62, 245–253. [Google Scholar]
45. Morelli, C.; Maranzano, P.; Otto, P. Spatiotemporal Clustering of GHGs Emissions in Europe: Exploring the Role of Spatial Component 2025. [Google Scholar]
46. Kang, Y.; Wu, K.; Gao, S.; Ng, I.; Rao, J.; Ye, S.; Zhang, F.; Fei, T. STICC: A Multivariate Spatial Clustering Method for Repeated Geographic Pattern Discovery with Consideration of Spatial Contiguity. International Journal of Geographical Information Science **2022**, 36, 1518–1549. [Google Scholar]

47. Turner, A.J.; Jacob, D.J.; Benmergui, J.; Brandman, J.; White, L.; Randles, C.A. Assessing the Capability of Different Satellite Observing Configurations to Resolve the Distribution of Methane Emissions at Kilometer Scales. Atmos. Chem. Phys. **2018**, 18, 8265–8278. [Google Scholar]
48. Saunois, M.; Stavert, A.R.; Poulter, B.; Bousquet, P.; Canadell, J.G.; Jackson, R.B.; Raymond, P.A.; Dlugokencky, E.J.; Houweling, S.; Patra, P.K.; Ciais, P. The Global Methane Budget 2000–2017. Earth Syst. Sci. Data **2020**, 12, 1561–1623. [Google Scholar]
49. Lan, X.; Basu, S.; Schwietzke, S.; Bruhwiler, L.M.P.; Dlugokencky, E.J.; Michel, S.E.; Sherwood, O.A.; Tans, P.P.; Thoning, K.; Etiope, G.; et al. Improved Constraints on Global Methane Emissions and Sinks Using $\delta^{13}$C-$CH_4$. Global Biogeochemical Cycles **2021**, 35, e2021GB007000. [Google Scholar]
50. Zhang, Z.; Zimmermann, N.E.; Calle, L.; Hurtt, G.; Chatterjee, A.; Poulter, B. Enhanced Response of Global Wetland Methane Emissions to the 2015–2016 El Niño-Southern Oscillation Event. Environ. Res. Lett. **2018**, 13, 074009. [Google Scholar]
51. Ciais, P.; Bastos, A.; Chevallier, F.; Lauerwald, R.; Poulter, B.; Canadell, J.G.; Hugelius, G.; Jackson, R.B.; Jain, A.; Jones, M.; et al. Definitions and Methods to Estimate Regional Land Carbon Fluxes for the Second Phase of the REgional Carbon Cycle Assessment and Processes Project (RECCAP-2). Geosci. Model Dev. **2022**, 15, 1289–1316. [Google Scholar]
52. Carranza, V.; Biggs, B.; Meyer, D.; Townsend-Small, A.; Thiruvenkatachari, R.R.; Venkatram, A.; Fischer, M.L.; Hopkins, F.M. Isotopic Signatures of Methane Emissions From Dairy Farms in California's San Joaquin Valley. JGR Biogeosciences **2022**, 127, e2021JG006675. [Google Scholar]
53. Ehret, T.; De Truchis, A.; Mazzolini, M.; Morel, J.-M.; Facciolo, G. Automatic Methane Plume Quantification Using Sentinel-2 Time Series. In Proceedings of the IGARSS 2022 - 2022 IEEE International Geoscience and Remote Sensing Symposium; IEEE: Kuala Lumpur, Malaysia, July 17 2022; pp. 1955–1958. [Google Scholar]
54. Dean, J.F.; Middelburg, J.J.; Röckmann, T.; Aerts, R.; Blauw, L.G.; Egger, M.; Jetten, M.S.M.; De Jong, A.E.E.; Meisel, O.H.; Rasigraf, O.; et al. Methane Feedbacks to the Global Climate System in a Warmer World. Reviews of Geophysics **2018**, 56, 207–250. [Google Scholar]
55. Turner, A.J.; Jacob, D.J.; Benmergui, J.; Wofsy, S.C.; Maasakkers, J.D.; Butz, A.; Hasekamp, O.; Biraud, S.C. A Large Increase in U.S. Methane Emissions over the Past Decade Inferred from Satellite Data and Surface Observations. Geophysical Research Letters **2016**, 43, 2218–2224. [Google Scholar]
56. Bi, H.; Neethirajan, S. Mapping Methane—The Impact of Dairy Farm Practices on Emissions Through Satellite Data and Machine Learning. Climate **2024**, 12, 223. [Google Scholar]
57. Johnson, K.A.; Johnson, D.E. Methane Emissions from Cattle. Journal of Animal Science **1995**, 73, 2483–2492, [Google Scholar]
58. Yu, X.; Millet, D.B.; Wells, K.C.; Griffis, T.J.; Chen, X.; Baker, J.M.; Conley, S.A.; Smith, M.L.; Gvakharia, A.; Kort, E.A.; Plant, G. Top-Down Constraints on Methane Point Source Emissions From Animal Agriculture and Waste Based on New Airborne Measurements in the U.S. Upper Midwest. JGR Biogeosciences **2020**, 125, e2019JG005429. [Google Scholar]
59. Pickett-Heaps, C.A.; Jacob, D.J.; Wecht, K.J.; Kort, E.A.; Wofsy, S.C.; Diskin, G.S.; Worthy, D.E.J.; Kaplan, J.O.; Bey, I.; Drevet, J. Magnitude and Seasonality of Wetland Methane Emissions from the Hudson Bay Lowlands (Canada). Atmos. Chem. Phys. **2011**, 11, 3773–3779. [Google Scholar]

60. Chen, Z.; Griffis, T.J.; Baker, J.M.; Millet, D.B.; Wood, J.D.; Dlugokencky, E.J.; Andrews, A.E.; Sweeney, C.; Hu, C.; Kolka, R.K. Source Partitioning of Methane Emissions and Its Seasonality in the U.S. Midwest. JGR Biogeosciences **2018**, 123, 646–659. [Google Scholar]
61. Bohn, T.J.; Melton, J.R.; Ito, A.; Kleinen, T.; Spahni, R.; Stocker, B.D.; Zhang, B.; Zhu, X.; Schroeder, R.; Glagolev, M.V.; et al. WETCHIMP-WSL: Intercomparison of Wetland Methane Emissions Models over West Siberia. Biogeosciences **2015**, 12, 3321–3349. [Google Scholar]
62. Bansal, S.; Post Van Der Burg, M.; Fern, R.R.; Jones, J.W.; Lo, R.; McKenna, O.P.; Tangen, B.A.; Zhang, Z.; Gleason, R.A. Large Increases in Methane Emissions Expected from North America's Largest Wetland Complex. Sci. Adv. **2023**, 9, eade1112. [Google Scholar]
63. Shen, L.; Zavala-Araiza, D.; Gautam, R.; Omara, M.; Scarpelli, T.; Sheng, J.; Sulprizio, M.P.; Zhuang, J.; Zhang, Y.; Qu, Z.; Lu, X. Unravelling a Large Methane Emission Discrepancy in Mexico Using Satellite Observations. Remote Sensing of Environment **2021**, 260, 112461. [Google Scholar]
64. VanderZaag, A.C.; Flesch, T.K.; Desjardins, R.L.; Baldé, H.; Wright, T. Measuring Methane Emissions from Two Dairy Farms: Seasonal and Manure-Management Effects. Agricultural and Forest Meteorology **2014**, 194, 259–267. [Google Scholar]
65. Islam, M.; Kim, S.-H.; Son, A.-R.; Ramos, S.C.; Jeong, C.-D.; Yu, Z.; Kang, S.H.; Cho, Y.-I.; Lee, S.-S.; Cho, K.-K.; Lee, S.S. Seasonal Influence on Rumen Microbiota, Rumen Fermentation, and Enteric Methane Emissions of Holstein and Jersey Steers under the Same Total Mixed Ration. Animals **2021**, 11, 1184. [Google Scholar]
66. Fouli, Y.; Hurlbert, M.; Kröbel, R. Greenhouse Gas Emissions from Canadian Agriculture: Estimates and Measurements. The School of Public Policy Publications **2021**, Vol. 14 No. 1 (2021). [Google Scholar]
67. Office of the Auditor General of Canada Agriculture and Climate Change Mitigation - Agriculture and Agri-Food Canada; Office of the Auditor General of Canada: Ottawa, Canada, 2024;
68. Kreft, C.; Finger, R.; Huber, R. Action- versus Results-based Policy Designs for Agricultural Climate Change Mitigation. Applied Eco Perspectives Pol **2024**, 46, 1010–1037. [Google Scholar]
69. Lobell, D.B.; Villoria, N.B. Reduced Benefits of Climate-Smart Agricultural Policies from Land-Use Spillovers. Nat Sustain **2023**, 6, 941–948. [Google Scholar]
70. Huang, W.; Griffis, T.J.; Hu, C.; Xiao, W.; Lee, X. Seasonal Variations of CH4 Emissions in the Yangtze River Delta Region of China Are Driven by Agricultural Activities. Adv. Atmos. Sci. **2021**, 38, 1537–1551. [Google Scholar]
71. Balasus, N.; Jacob, D.J.; Lorente, A.; Maasakkers, J.D.; Parker, R.J.; Boesch, H.; Chen, Z.; Kelp, M.M.; Nesser, H.; Varon, D.J. A Blended TROPOMI+GOSAT Satellite Data Product for Atmospheric Methane Using Machine Learning to Correct Retrieval Biases. Atmos. Meas. Tech. **2023**, 16, 3787–3807. [Google Scholar]
72. Williams, J.P.; Ars, S.; Vogel, F.; Regehr, A.; Kang, M. Differentiating and Mitigating Methane Emissions from Fugitive Leaks from Natural Gas Distribution, Historic Landfills, and Manholes in Montréal, Canada. Environ. Sci. Technol. **2022**, 56, 16686–16694. [Google Scholar]
73. Ishizawa, M.; Chan, D.; Worthy, D.; Chan, E.; Vogel, F.; Melton, J.R.; Arora, V.K. Estimation of Canada's Methane Emissions: Inverse Modelling Analysis Using the Environment and Climate Change Canada (ECCC) Measurement Network. Atmos. Chem. Phys. **2024**, 24, 10013–10038. [Google Scholar]

74. Li, H.Z.; Seymour, S.P.; MacKay, K.; Wang, J.S.; Warren, J.; Guanter, L.; Zavala-Araiza, D.; Smith, M.L.; Xie, D. Direct Measurements of Methane Emissions from Key Facilities in Alberta's Oil and Gas Supply Chain. Science of The Total Environment **2024**, 912, 169645. [Google Scholar]